\documentclass[
10pt,
aps,
pra,
twocolumn,
floatfix,
nofootinbib,
superscriptaddress,
longbibliography
]{revtex4-2}

\usepackage{bm,graphicx,mathrsfs,amsmath,amssymb,mathtools,makecell,bbm,amsthm,dsfont,color,times,txfonts,nicefrac,framed,enumitem,tikz,physics,wrapfig,amsfonts,nicematrix,xargs}

\usepackage{float}
\usepackage{subcaption}
\usepackage{algorithm}
\usepackage{algpseudocode}
\usepackage{multirow} 
\usepackage{nicefrac}       
\usepackage{verbatim}       
\usepackage[dvipsnames,svgnames,x11names]{xcolor}
\definecolor{refscolor}{RGB}{190,120,230}
\usepackage{booktabs}       

\usepackage[colorlinks=true,linkcolor=refscolor,citecolor=refscolor]{hyperref}
\hypersetup{urlcolor=refscolor}

\usepackage[breakable]{tcolorbox}
\newtcolorbox{mybox}[1]{breakable,%
  breakable,%
  colback=boxcolor!5!white,%
  colframe=boxcolor!75!black,%
  fonttitle=\bfseries,%
  title=#1%
}

\usepackage{fancyhdr}
\newcommand{\xhdr}[1]{{\noindent\bfseries #1}.}

\begin{document}
\title{Inductive Graph Representation Learning with Quantum Graph Neural Networks}

\author{Arthur M. Faria}
    \affiliation{Pasqal NL, Fred. Roeskestraat 100, 1076 ED Amsterdam, Netherlands}
    \email{arthur.faria2@gmail.com}
\author{Ignacio F. Graña}
    \affiliation{Pasqal NL, Fred. Roeskestraat 100, 1076 ED Amsterdam, Netherlands}
    \email{ignacio.fernandez-grana@pasqal.com}    
\author{Savvas Varsamopoulos}
    \affiliation{Pasqal NL, Fred. Roeskestraat 100, 1076 ED Amsterdam, Netherlands}
    \email{savvas.varsamopoulos@pasqal.com}    

\date{\today}

\begin{abstract}
Quantum Graph Neural Networks (QGNNs) offer a promising approach to combining quantum computing with graph-structured data processing. While classical Graph Neural Networks (GNNs) are scalable and robust, existing QGNNs often lack flexibility due to graph-specific quantum circuit designs, limiting their applicability to diverse real-world problems. To address this, we propose a versatile QGNN framework inspired by GraphSAGE, using quantum models as aggregators to generate quantum-native node embeddings. We integrate inductive representation learning techniques with parameterized quantum convolutional and pooling layers, bridging classical and quantum paradigms. The convolutional layer is flexible, allowing tailored designs for specific tasks. Benchmarked on a node regression task with the QM9 dataset, our framework, using a single minimal circuit for all aggregation steps, handles molecules with varying numbers of atoms without changing qubits or circuit architecture. While classical GNNs achieve higher training performance, our quantum approach remains competitive and often shows enhanced generalization as molecular complexity increases. We also observe faster learning in early training epochs. To mitigate trainability limitations of a single-circuit setup, we extend the framework with multiple quantum aggregators on QM9. Assigning distinct circuits to each hop substantially improves training performance across all cases. Additionally, we numerically demonstrate the absence of barren plateaus as qubit numbers increase, suggesting that the proposed model remains trainable to larger, more complex graph-based problems.
\end{abstract}

\maketitle
\thispagestyle{plain}
\section{Introduction}\label{intro}
Graphs provide a versatile framework for modeling complex relationships, with applications ranging from molecular structures to social networks. Unlike traditional machine learning models such as convolutional neural networks (CNNs) and feedforward neural networks (NNs), graph neural networks (GNNs) are specifically designed to exploit the relational structure inherent in graph data. This unique capability enables GNNs to excel in tasks such as node classification, link prediction, and graph regression, where understanding the relationships between entities is as critical as understanding the entities themselves. As a result, GNNs have found widespread use across diverse domains, including social network analysis~\cite{borisyuk_2024_lignn, fan_2019_graph}, recommendation systems~\cite{de_2024_personalized, jain_2019_food}, drug discovery~\cite{stokes_2020_deep}, traffic prediction~\cite{derrow_2021_eta}, and polypharmacy side effects~\cite{kondor_modeling_polypharmacy_2023, zitnik_2018_modeling}.

While classical GNNs have achieved remarkable success, the emergence of quantum computing introduces transformative potential for machine learning. Quantum Machine Learning (QML) harnesses principles of quantum mechanics, such as entanglement and superposition have been shown to yield significant speedups over classical computing for specific tasks~\cite{arute2019quantum, google2025, cerezo_2022_challenges, coles_2021_seeking, perdomo_2018, ciliberto_2018, liu_2021}. In the context of graph-based problems, Quantum Graph Neural Networks (QGNNs) hold significant potential for addressing generalization limitations of classical models on graph-structured data~\cite{liao2021}, particularly across graphs of varying size and topology, as well as, more broadly, scalability challenges~\cite{ma2022}.

Early efforts to develop QGNNs have  demonstrated promise by extending  classical frameworks as Graph Convolutional Networks (GCNs)~\cite{welling_GCN_2017} to the quantum domain via digital quantum circuits~\cite{hu_qgnn_2022, yidong_qgnn_2024, ryu_qgnn_2023, zheng_qgnn_2021, zheng_qgnn_2024}. However, despite these advances, the existing QGNN frameworks still face significant limitations in scalability and in their ability to generalize to unseen data. This is largely because their quantum circuits are tailored to fixed graph structures, preventing them from adapting to the diversity of real-world graph datasets. As a result, current quantum approaches struggle to meet the demands of practical applications, where adaptability, scalability, and strong generalization are essential.

\xhdr{Present Work} To address these challenges, we propose a versatile QGNN framework for inductive node embedding, extending into the quantum realm the classical GraphSAGE approach~\cite{leskovec_inductive_representation_2018}. This framework samples uniformly neighborhood nodes and uses quantum circuits as aggregators to efficiently combine their features in an inductive manner, enabling both scalability and generalization to unseen data. The circuit is composed of parameterized quantum convolutional and pooling layers, inspired by the Quantum Convolutional Neural Network (QCNN) architecture~\cite{lukin_qcnn_2019}. 

To rigorously evaluate the performance of our QGNN framework, we conduct node regression experiments on the QM9 dataset~\cite{pyg_code, moleculeNet_2017}. Molecules are represented as graphs, in which the nodes are corresponding to the atoms and edges to the bonds. Our results demonstrate that the quantum model performs well on the sample data, achieving strong performance metrics. However, when using a single quantum circuit reused across all aggregation steps, classical GNNs consistently outperform the quantum model during training, whereas the quantum model exhibits stronger generalization, particularly as molecular size increases (i.e., for molecules with a larger number of atoms).
This highlights the robustness of our QGNN framework in maintaining consistent generalization across molecules of varying sizes, despite relying on a minimal model. The aforementioned trainability limitations of the single-circuit setup are addressed by introducing multiple independent quantum aggregators, leading to higher training performance of the quantum approach. Furthermore, we investigate the scalability of the QGNN framework with respect to an increasing number of qubits. This is achieved by addressing the barren plateau problem~\cite{anschuetz_2022_beyond, ragone_2023_lie, larocca_2024_review, arrasmith_2021_equivalence}, where gradients vanish exponentially, hindering optimization. Leveraging previous results~\cite{ragone_2023_lie} and using random input node features, we numerically show that the proposed QGNNs do not exhibit barren plateaus since the variance of the loss function partial derivatives scales sub-exponentially with the number of qubits. This suggests that our framework remains trainable at scale, motivating its use for further investigations on more complex graph-based data.

In summary, our work makes four key contributions:  

    \vspace{-1mm}
\begin{enumerate}  
    \item We introduce a novel QGNN framework that integrates standard graph sampling strategies with quantum convolutional and quantum pooling layers.  
    \vspace{-2mm}

    \item We evaluate our approach on QM9 using a single minimal QGCN circuit, reused at every aggregation step, to show that it can handle graphs with varying numbers of nodes without changing the number of input qubits.  
    \vspace{-2mm}

    \item We improve the trainability of QGNNs by extending the framework incorporate multiple quantum aggregators.
    \vspace{-2mm}

    \item We numerically investigate the QGNN framework's scalability by confirming, via the variance of the loss gradients, the absence of barren plateaus as the number of qubits increases.
\end{enumerate} 

The paper is structured as follows: Section~\ref{sec:data} introduces the problem dataset and explains the node regression task we aim to investigate in this work. Section~\ref{sec:related_work} reviews prior work on QGNNs, identifying their limitations and proposing how our framework addresses these challenges. Section~\ref{sec:gnn} provides a foundational overview of classical GNNs, with a particular focus on the GraphSAGE approach. Section~\ref{sec:qgnn} details our proposed QGNN framework, including the design of the quantum circuit and a discussion of its advantages and potential benefits. Section~\ref{sec:results} presents the numerical results for both classical and quantum models benchmarked on the QM9 dataset, comparing their performance. We also numerically confirm the absence of barren plateaus in our QGNN. Finally, Section~\ref{sec:conclusions} discusses the results and provides concluding remarks.

\section{Molecular Data (QM9)}\label{sec:data}
The QM9 dataset~\cite{pyg_code, moleculeNet_2017} is a standard benchmark in computational chemistry, comprising small organic molecules built from combinations of nine heavy atoms, including carbon, oxygen, nitrogen, and fluorine. It provides geometric, energetic, electronic, and thermodynamic properties calculated using density functional theory (DFT), including atomization energies, dipole moments, and HOMO-LUMO gaps (Highest-Lowest Unoccupied Molecular Orbital). In particular, the LUMO and HOMO energies are key indicators of a molecule's reactivity and stability. These properties are inherently molecular, describing the molecule as a whole rather than individual atoms.

Specifically, we focus on a node regression task, i.e., using atomic properties such as node features to predict molecular properties. The dataset provides 11 atomic features, but we use only 7 of them: atomic number, chirality, degree, formal charge, radical electrons, hybridization, and scaled mass. The remaining four (aromaticity, hydrogen count, ring membership, and valence) are excluded since they are consistently zero in the selected samples used in this work.
In this way, node regression leverages local atomic information to infer global molecular behavior. Essentially, in this investigation, each molecule is represented as a graph, where nodes represents atoms, and edges only indicate connectivity between atoms.  For each molecule in the QM9 dataset, we extract atom features and the adjacency matrix. However, only the atom features are used as classical input to the model, while the adjacency matrix specifies the set of atom features to be used as input.

\section{Related Work}\label{sec:related_work}
Quantum Graph Neural Networks have emerged as a promising approach to leverage quantum computing for graph-based learning tasks. Verdon et al.~\cite{verdon_2019_quantumgraph} introduced one of the first analog
QGNN frameworks, leveraging Hamiltonian evolution through global unitary dynamics for tasks such as Hamiltonian learning, entanglement generation, and graph isomorphism classification. Building on this, Henry et al.~\cite{thabet_2024} extended the concept by applying quantum kernel methods to classification tasks.

In contrast to analog approaches, a few works have shifted toward digital implementations using variational circuits. For instance, Beer et al.~\cite{beer_2021_quantum} developed quantum neural networks for graph-structured quantum data, while Skolik et al.~\cite{skolik_2023_equivariant} introduced QAOA-type circuits designed for weighted graphs.  Hybrid quantum–classical approaches have also emerged as possible alternatives to constraints posed by quantum circuits. Ai et al.~\cite{xing_qgnn_2024} proposed a framework that decomposes large graphs into smaller subgraphs, and Zhou et al.~\cite{kaixiong_qgnn_2022} combined classical clustering with variational circuits. Expanding on hybrid approaches, Tuysuz et al.~\cite{tuysuz_2021_hybrid} demonstrated a quantum-classical GNN applied to the reconstruction of the particle track, further developing hybrid methods. 

Recent progress has been made in fully digital QGNNs, yet major challenges remain. Zheng et al.~\cite{zheng_qgnn_2021,zheng_qgnn_2024} introduced QGCN architectures with convolution and pooling layers, addressing barren plateaus and reporting strong performance. However, their design encodes the full graph—adjacency, nodes, and edges—within a single quantum circuit, causing circuit width and depth to scale rapidly with graph size. Liao et al.~\cite{yidong_qgnn_2024} face similar overheads due to large circuit constructions and lack numerical benchmarks. Hu et al.~\cite{hu_qgnn_2022} design an edge-based QGCN for semi-supervised node classification, where circuit depth can increase rapidly with the number of edges, potentially resulting in large circuits and training bottlenecks for graphs with particular symmetries. All these approaches rely on amplitude encoding. Ryu et al.~\cite{ryu_qgnn_2023} employ equivariantly diagonalizable unitary quantum graph circuits (EDU-QGCs) for node regression in molecules, but the atom-to-qubit mapping becomes costly for larger molecules. Finally, Mernyei et al.~\cite{mernyei_2022_equivariant} evaluate equivariant quantum graph circuits only on small synthetic datasets, leaving scalability to more complex graphs uncertain.

Conversely, our QGNN framework adapts GraphSAGE~\cite{leskovec_inductive_representation_2018} for quantum computing, providing a flexible inductive learning architecture that lays the foundation for improved generalization across graphs of varying size and structure. By constructing the graph's computational representation instead of directly encoding its adjacency matrix into the quantum circuit, we preserve topological features without introducing unnecessary quantum circuit overhead. The circuit integrated into this framework is designed to incorporate flexible convolutional and pooling layers, enabling users to design various Ansätze. Importantly, the number of qubits depends solely on the node feature dimension, keeping the circuit architecture invariant to the number of sampled nodes and edges. Combined with that, we employ a nonlinear quantum feature map, instead of amplitude encoding, which offers rich expressivity for reasonably shallow circuit depths~\cite{kyriienko_2021_solving}. By addressing the limitations outlined above, our framework offers a solid foundation for advancing quantum graph machine learning, including potential extensions of classical heterogeneous-graph implementation to the quantum domain, as in Refs.~\cite{kondor_modeling_polypharmacy_2023, zitnik_2018_modeling, leskovec_embedding_2019}.

\section{Classical GNN: GraphSAGE}\label{sec:gnn}
GNN handle graph-structured data, where nodes represent entities and edges capture relationships between them. Unlike traditional neural networks that operate on grid-like data structures, GNNs manage flexible, interconnected structures, making them suitable for diverse applications, including social networks, molecular modeling, and knowledge graphs.

\begin{figure}[H]
    \centering
    \begin{subfigure}{0.38\textwidth}
        \includegraphics[width=\textwidth]{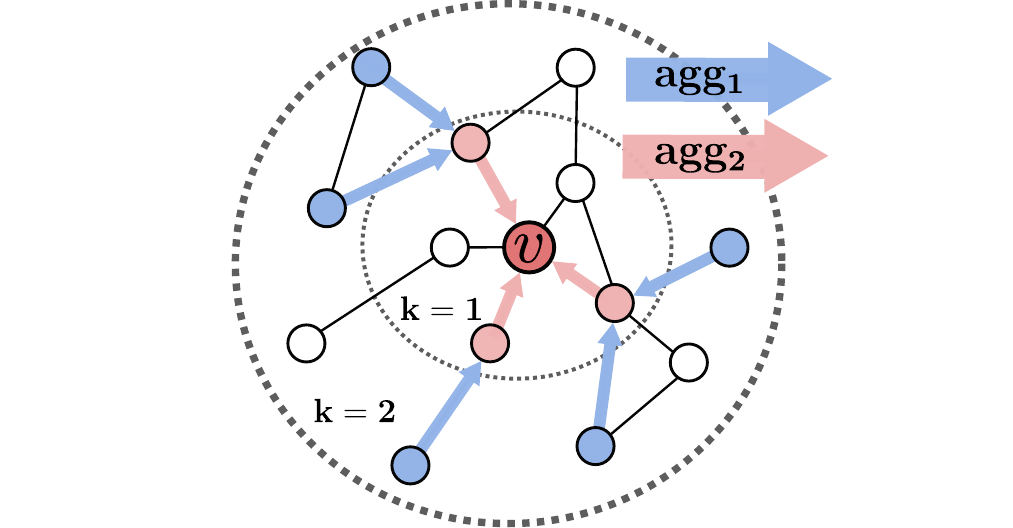}
        \caption{Message passing and aggregation mechanisms.}
        \label{fig:aggr_comp}
    \end{subfigure}
    \begin{subfigure}{0.45\textwidth}
    \vspace{1mm}
        \includegraphics[width=\textwidth]{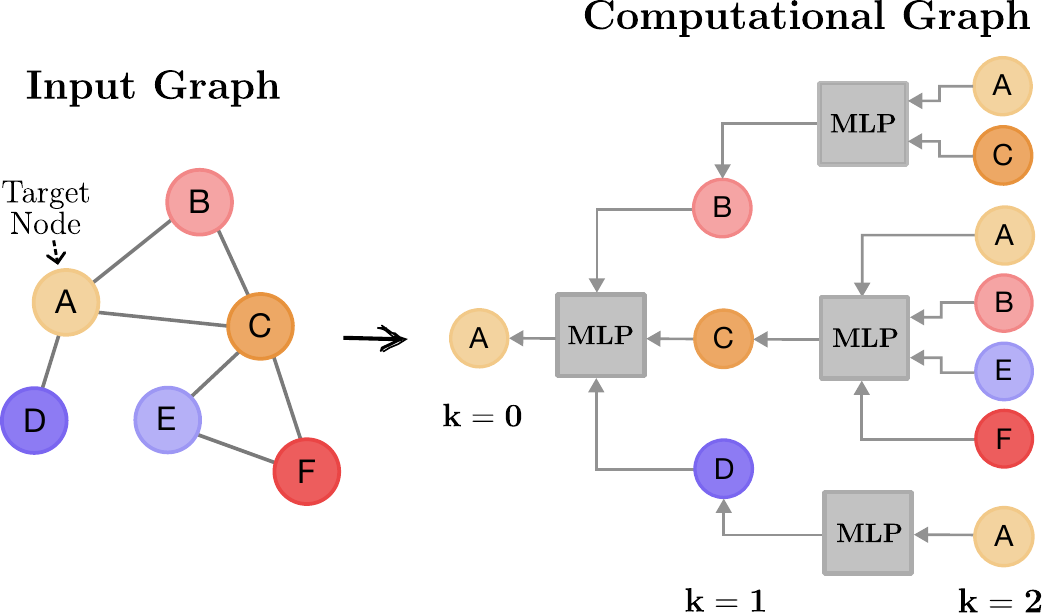}
        \caption{Computational graph scheme.}
        \label{fig:comp_graph}
    \end{subfigure}
    \vspace{-2mm}
    \caption{Illustration of the GraphSAGE framework. (a) Aggregation steps: agg1 and agg2, generate the message which is then passed to the target node $v$ for $k=2$ hops. (b) Transformation of an input graph (left) into a computation graph (right), where nodes represent feature embeddings and edges indicate information flow. Gray boxes denote Multi-Layer Perceptrons (MLPs), which aggregate neighboring embeddings through modular transformations, enabling accurate node representations.}
    \label{fig:graphsage}
\end{figure}

GraphSAGE is a general inductive framework that leverages node feature information to efficiently generate node embeddings. At its core are the sampling and aggregation mechanisms.  Using the input graph's adjacency matrix, the computation graph representation is constructed, preserving local graph structures. Fig.~\ref{fig:graphsage} provides a schematic representation of the method, where in particular the message generation and passing are illustrated in Fig.~\ref{fig:aggr_comp}.

For a given depth or "hop" of the computational graph, the framework samples a fixed number of neighbors. Embeddings are dynamically generated via a modular aggregation function, producing local neighborhood embeddings for target nodes. Additionally, concatenation can be performed to incorporate the target node's features with the aggregated neighborhood information. By stacking multiple hops, the target node embedding incorporates information from increasingly larger neighborhoods, capturing multi-hop structural information. This captures neighborhood topology, enabling scalability to large graphs and generalization to unseen data.

We now delve into the mathematical fundamentals of GraphSAGE, with a focus on the embedding generation and its message after the sampling and aggregation steps

\vspace{-6mm}

\subsection{Inner Workings of GraphSAGE}
In the GraphSAGE framework, the GNN generates node embeddings through two main stages: neighborhood sampling and feature aggregation, where the aggregation is performed by an MLP-based aggregator. In particular, the MLP are defined by an inductive variant of the GCN. For brevity, we refer to this aggregator simply as GCN throughout the text. 

In the layer $l=0$, the embedding of the target node $v$ is $\mathbf{h}_v^{(0)} = \mathbf{x}_v$, where $\mathbf{x}_v$ represents the input features of $v$. Based on the computational graph, the sampling step uniformly selects the neighbors $u$ of node $v$ and retrieves their feature vectors. The message (\textsc{Msg}) generated at the GCN layer $l$ for the neighbor $u$ of $v$ reads
\begin{align}
    \mathbf{m}_u^{(l)} = \textsc{Msg}^{(l)} \left( \{\mathbf{h}_u^{(l-1)}, \forall u \in \mathcal{N}(v)\} \right)
\end{align}
\noindent where $\mathcal{N}(v)$ represents the set of neighbors of node $v$. The embeddings $\mathbf{h}_u^{(l-1)}$ correspond to the representations of neighbor $u$ from the previous layer, $l-1$. In the message-passing process, learnable parameters $\mathbf{W}^{(l)}$ (weights) and $\mathbf{B}^{(l)}$ (biases) are assigned at layer $l$ to the neighboring nodes $u$ and the target node $v$, respectively. In the aggregation step, neighbor messages are combined as follows
\begin{align}
    \mathbf{h}_v^{(l)} = \textsc{Agg}^{(l)} \left( \{\mathbf{m}_u^{(l)}, u \in \mathcal{N}(v)\}, \mathbf{m}_v^{(l)} \right),
\end{align}
\noindent being the aggregation $(\textsc{Agg})$ originally defined as the average of the neighboring nodes' embeddings.  In contrast, GraphSAGE introduces an intermediate step, where $\mathbf{h}_v^{(l)}$ is not updated automatically. In fact, it first enables a flexible aggregation process to generate the aggregated embedding of neighboring nodes $\mathbf{h}_{\mathcal{N}(v)}^{(l)}$, typically expressed as 
\begin{align}
    \mathbf{h}_{\mathcal{N}(v)}^{(l)} = \textsc{Agg}^{(l)} \left( \{\mathbf{h}_u^{(l-1)}, \forall u \in \mathcal{N}(v)\} \right).
\end{align}
\noindent where $\textsc{Agg}$ can be any general vector function. Common aggregation methods include $\textsc{Mean}$~\cite{welling_GCN_2017}, $\textsc{LSTM}$~\cite{hochreiter_1997_long}, or $\textsc{Pooling}$~\cite{qi_2016_pointnet}. In this work, we choose the $\textsc{Mean}$ function due to its simplicity and permutation invariance, i.e., $\mathbf{h}_{\mathcal{N}(v)}^{(l)} = \textsc{mean}(\{\mathbf{h}_u^{(l-1)}, \forall u \in {\mathcal{N}(v)}\}$. This type of aggregator computes the element-wise mean of the neighbor embeddings $\mathbf{h}_u^{(l-1)}$. GraphSAGE also performs concatenation of the target node's previous-layer embedding $\mathbf{h}^{(l-1)}_v$ with $\mathbf{h}^{(l)}_{\mathcal{N}(v)}$, which is not immediately done in~\cite{welling_GCN_2017}. The final embedding of $v$ reads
\begin{align}
    \mathbf{h}_v^{(l)} = \sigma \left( \mathbf{W}^{(l)} \cdot \left[ 
    \mathbf{h}_{\mathcal{N}(v)}^{(l)} \, \| \, \mathbf{h}_v^{(l-1)} 
    \right] \right)
    \label{eq:concat}
\end{align}
\noindent where $[\cdot \| \cdot]$ denotes concatenation and $\sigma$ is a nonlinear activation function that enhances expressivity.  Alternatively, some GNN architectures refine this process by integrating attention mechanisms, which dynamically adjust the influence of each neighbor based on its relevance~\cite{petar_gats_2018}. 

In the layer $L$, the final node embedding, $\mathbf{h}_v^{(L)}$ encodes both the target node’s intrinsic features and its structural relationships within the graph. Hence, GraphSAGE effectively captures the structural information of the graph by learning the role of a node within it. Building upon these foundations, we now extend this framework to the quantum domain, introducing our QGNNs.

\section{Proposed QGNN Framework}\label{sec:qgnn}
In this section, we present our QGNN framework, which integrates quantum circuits with GraphSAGE techniques to generate node embeddings. The QGNN framework comprises three key components: (1) the updated target node embedding is generated by aggregating and concatenating messages from neighboring nodes. (2) A feature map that maps classical features into qubit states. (3) A QGCN circuit that acts as the quantum modular aggregator, combining the embeddings of connected nodes.

\begin{figure}[H]
    \centering
    \includegraphics[width=0.33\textwidth]{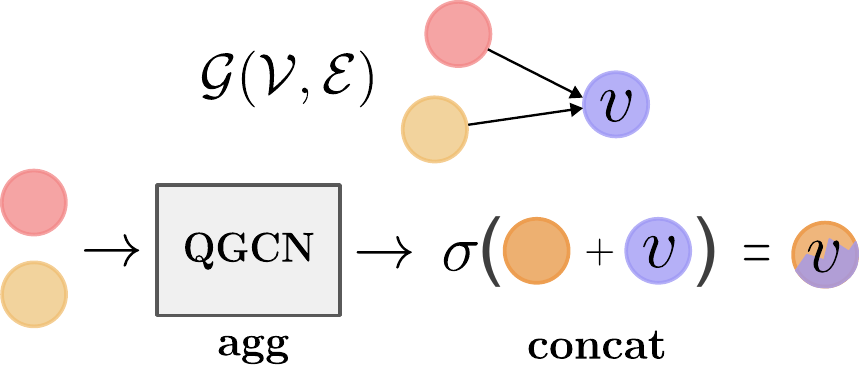}
    \caption{Schematic representation of the proposed QGNN framework. Embedding generation process: based on the computational graph of $\mathcal{G} = (\mathcal{V}, \mathcal{E})$, node features of connected nodes to $v$ (red and yellow) are sampled uniformly and input into the QGCN. The circuit aggregates these features to create a single, modular neighboring-node embedding (orange) $\mathbf{h}_{\mathcal{N}(v)}^{(L)}(\bm{\theta}_v)$. This is concatenated classically with the target node’s self-embedding (purple) through a simple vector operation and then processed by a non-linear function $\sigma$ to yield the updated embedding $\mathbf{h}_v^{(L)}(\bm{\theta}_v)$.}
    \label{fig:framework}
\end{figure}

The first component, inspired by GraphSAGE, utilizes the adjacency matrix of the input graph to construct its computation graph representation, preserving local graph structures. This enables the identification of connected nodes $u$ of $v$, i.e., $\{u \in \mathcal{V} : (u, v) \in \mathcal{E}\}$, whose features are sampled uniformly and input into an MLP (see Fig.~\ref{fig:comp_graph}). The second component initializes the qubit states, while the third employs QGCNs to perform GraphSAGE-style aggregation, replacing the classical MLPs traditionally used in the message-passing phase. This approach leverages quantum mechanics to process local classical node features. Together, these elements form our QGNN architecture for node-level tasks. Fig.~\ref{fig:framework} illustrates the overall framework.

\vspace{-2mm}

\subsection{Quantum feature map}
The classical input feature of node $u$, denoted by $\mathbf{x}_u$, is passed into the QGCN through a feature map. The feature map $\mathcal{F}$ is realized by a unitary $U_{x}$ acting on the initial $n$-qubit state $|0\rangle^{\otimes N}$, spanning the Hilbert space $\mathcal{H}$. Concretely,
\begin{align}
    \mathbf{x}_u &\mapsto |\mathcal{F}(\mathbf{x}_u)\rangle = U_{x}|0\rangle^{\otimes N},
\end{align}
Hence, the initial state of $u$,  $\rho_u$, becomes
\begin{align}
    \rho_u = \lvert \mathcal{F}(\mathbf{x}_u) \rangle \langle \mathcal{F}(\mathbf{x}_u) \rvert.
    \label{eq:qubit_init}
\end{align}
In this particular work, we employ the Fourier quantum feature map to improve expressivity, as it effectively maps highly nonlinear classical functions into qubit states~\cite{kyriienko_2021_solving, schuld_2021}. For a set of $N$ neighbors of node $v$, the overall input state becomes
\begin{align}
    \rho_{\mathrm{in}}= \bigotimes_{w=1}^N \, \rho_{w},
    \label{eq:initial_state}
\end{align}
Eq.~\eqref{eq:initial_state} implies that each state $\rho_w$ has the same dimension, since the number of qubits defining it depends solely on the dimension of $\mathbf{x}_u$ (see Eq.~\eqref{eq:qubit_init}), which is assumed to be identical for all vectors. By feeding node features into the circuit sequentially, the architecture remains independent of the number of sampled nodes and edges. The number of qubits may be increased to allow richer encodings, but the minimum must match the feature dimension. Now, let us describe the Ansatz.

\subsection{The QGCN circuitry}
Inspired by previous QCNN designs~\cite{lukin_qcnn_2019}, the QGCN circuit also alternates between quantum convolutional and quantum pooling layers and defines the local Ansatz. 

Convolutional layers capture local graph structure by applying learnable two-qubit operations that generate entanglement between neighbouring qubits. This produces quantum states that encode the relevant local features. Conversely, pooling layers reduce the dimensionality of the quantum state while retaining its essential features, thereby lowering computational cost without losing key information. In this way, pooling yields a coarser quantum representation that complements the local state transformations performed by the convolutional layer. 

The QGCN circuit consists of a composition of $L$ quantum channel layers, defined as
\begin{align}
     \Phi_{\bm{\theta}_v} =
     \bigcirc_{l=1}^{L} \left(\mathrm{P}_{l} \circ \mathrm{C}_{l}^{\bm{\theta}^{l}_v}\right),
     \label{eq:qgcn-map}
\end{align}
\noindent where $\bm{\theta}_v := \{\bm{\theta}^{l}_v\}_{l=1}^L$. We denote a single function composition by $\circ$, while $\bigcirc_{l=1}^L$ refers to the sequential composition of $L$ functions. Importantly, each QGCN layer $l$ is built from a quantum convolution layer $\mathrm{C}_l^{\bm{\theta}^{l}_v}$, parametrized by $\bm{\theta}_v^l$, followed by a quantum pooling layer $\mathrm{P}_l$. Since only the convolutional layers carry parameters, they define the Ansatz. Pooling layers could also be parametrized, e.g. $\mathrm{P}_l^{\bm{\lambda}_l}$, to allow more flexible pooling operations, but we keep the architecture minimal to emulate the quantum analogue of the classical mean pooling. The alternating sequence of convolution and pooling layers progressively transforms and reduces the quantum state, beginning with $\mathrm{C}_1$ and ending with $\mathrm{P}_L$.

\begin{figure}[H]
    \centering
    \includegraphics[width=0.31\textwidth]{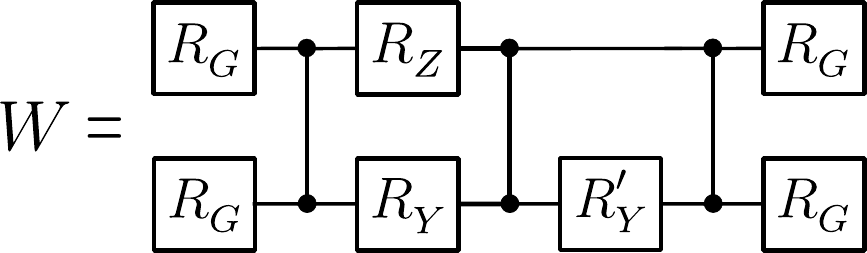}
    \caption{Schematic representation of the second convolutional unitary $W$, acting on two neighboring qubits. Each layer comprises of CZ gates applied to alternating qubit pairs, enclosed by parameterized general single-qubit rotations $R_G$. The single qubit gates: $R_Z, R_Y$, and, $R'_Y$ also are parametrized with different tunable angles.}
    \label{fig:ansatz}
\end{figure}

The convolutional layer $\mathrm{C}_l^{\bm{\theta}_v^l}$: $\mathcal{S}(\mathcal{H}_l) \rightarrow \mathcal{S}(\mathcal{H}_l)$ preserves the size of the quantum register. It reads
\begin{align}
    \mathrm{C}_l^{\bm{\theta}_v^l}(\cdot) =
    \scalebox{1}{$\bigcirc$}_{j=1}^{r} \left(\bigotimes_{(i,i+1)  \in \mathrm{S}(j)} W_l^{(i,i+1)}\left(\bm{\theta}_v^l\right)\right)(\cdot),
    \label{eq:conv_layers}
\end{align}
Within each convolutional layer, which acts on an even number of qubits, the unitary $W_l$—referred to as the convolutional cell—operates on pairs of neighboring qubits $(i, i+1)$, as depicted in Fig.~\ref{fig:ansatz}. For $n_l$ qubits in layer $l$, the operator $W_l$ acts on adjacent qubit pairs in an alternating pattern: for even values of $j$, it targets the pairs $(0,1), (2,3), \dots$, while for odd $j$, it instead acts on $(1,2), (3,4), \dots$. Formally, the set of qubit pairs involved are given by $S(j) = \{(i, i+1) \,\mid\, i \equiv j \pmod{2},\, 0 \leq i \leq n_l-2 \}$. The parameter $r$ specifies the depth of each convolutional layer $\mathrm{C}_l^{\bm{\theta}_v^l}(\cdot)$.  For example, in a two-layer QGCN architecture, the depths of $\mathcal{C}_{1}$ and  $\mathcal{C}_{2}$ are denoted as $r_{1}$ and $r_{2}$, respectively. This is $r = [r_1,r_2]$. 

Following Ref.~\cite{vatan_2004_optimal}, $W_l$ is generally decomposed into 3 CZ gates, 3 single-qubit rotations, and 4 general $R^G$ gates, where
\begin{align}
    R^G(\bm{\theta}_v^l) = e^{-iX\theta_1^{l}/2}e^{-iZ\theta_2^{l}/2}e^{-iX\theta_3^{l}/2}.
\end{align}
The unitary $W$ is defined as in Fig.~\ref{fig:ansatz}. The convolutional cell $W$ is customizable, and its definition uniquely determines the convolutional layers. In particular, QGCN layers can be designed to preserve data symmetries~\cite{larocca_2022_group, chinzei_2023, cerezo_2022_theory_equivariant}, making them particularly effective for symmetry-equivariant tasks. The circuit can also be extended to include fault-tolerant operations on QEC codes~\cite{lukin_qcnn_2019, zeng_2016}. 

\vspace{3mm}

\begin{figure}[H]
    \centering
    \vspace{-6mm}
    \includegraphics[width=0.48\textwidth]{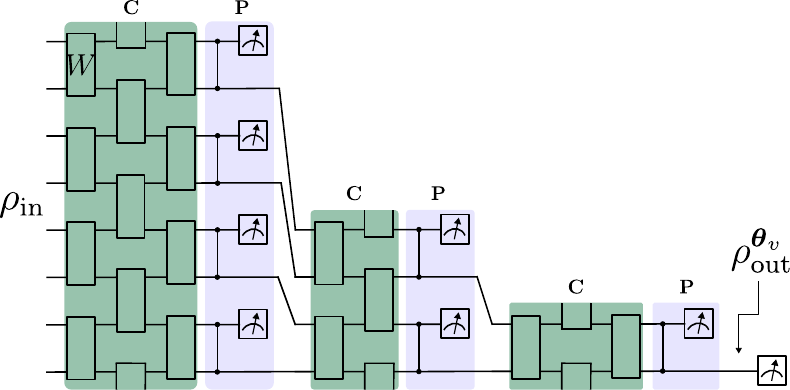}
    \caption{Schematic representation of the QGCN circuit. The QGCN takes as input an $n$-qubit quantum state $\rho_{\text{in}}$ and processes it through a sequence of $L$ alternating convolutional (C) and pooling (P) layers. In each convolutional layer, parameterized quantum gates perform unitary transformations $W$ on pairs of neighboring qubits. The pooling layers then reduce the number of qubits by tracing out half of them between layers. After the last pooling layer, the circuit produces the output state $\rho_{\mathrm{out}}^{\bm{\theta}_v}$.}
    \label{fig:qgnn_circuit}
\end{figure}
\vspace{2mm}
Conversely, the pooling layer $\mathrm{P}_l$: $\mathcal{S}(\mathcal{H}_l) \rightarrow \mathcal{S}(\mathcal{H}_{l+1})$ reduces the size of the quantum register by tracing out half of the qubits, such that $\dim(\mathcal{H}_{l+1}) < \dim(\mathcal{H}_l)$. It reads
\begin{align}\label{eq:pooling-layer}
    \mathrm{P}_{l} = \Tr_{i}[(\cdot)].
\end{align}
Here, $i$ denotes the qubits traced out at layer $l$, which halves the quantum register at each pooling step, i.e., $i = \{ j \in \{0,1,\dots,n_l-1\} \,|\, j \equiv 0 \pmod{2} \}$. Typically, the layer is implemented by measuring a subset of qubits, whose classical outcomes condition the application of controlled Pauli gates on their neighbors. Our design follows, however, the principle of deferred measurement~\cite{nielsen}. Hence, the pooling layers first employ quantum controlled operations,  deferring the (partial) measurements to the end. 

At the final QGCN layer $L$, the modular output state produced by the circuit, $\rho_{\mathrm{out}}^{\bm{\theta}_v}$, reads 
\begin{align}
    \rho_{\mathrm{out}}^{\bm{\theta}_v}&= \Phi_{\bm{\theta}_v}(\rho_{\mathrm{in}})=\bigotimes_{w=1}^N \, \Phi_{\bm{\theta}_v} (\rho_{w})
\end{align}
The complete QGCN circuit is illustrated in Fig.~\ref{fig:qgnn_circuit}. Note that the CZ gates employed to entangle qubit states, both in the pooling layers and in the convolutional cell $W_l$ definition, can be easily replaced by other two-qubit operations, such as CNOTs.

\subsection{Embedding generation and model training}

After employing the trainable FM to initialize the qubit states and using the QGCN as the quantum aggregator, the classical embedding for the neighbors $u$ of $v$ is obtained through measurements of an observable ${\mathcal{O}}$, expressed as
\begin{align}
    \mathbf{h}_{\mathcal{N}(v)}^{(L)}(\bm{\theta}_v) &=
    \Tr[\rho_{\mathrm{out}}^{\bm{\theta}_v} {\mathcal{O}}]= \langle {\mathcal{O}}\rangle_{\rho_{\mathrm{out}}^{ \bm{\theta}_v}} .
\end{align}
Note that the embedding is generated by the same QGCN regardless of the number of neighboring nodes $u$. It can be further concatenated with the self-embedding of node $v$ to yield its updated representation as
\begin{align}
    \mathbf{h}_v^{(L)}(\bm{\theta}_v) = \sigma \left(\left[ 
    \mathbf{h}_{\mathcal{N}(v)}^{(L)}(\bm{\theta}_v) \, \| \, \mathbf{h}_v^{(1)} \right] \right).
    \label{eq:concat2}
\end{align}
As the QGCN circuit has $L$ layers, the initial embedding $\mathbf{h}_v^{(1)}$ can be only concatenated after the quantum measurement. In addition, the concatenation operation used here is a simple vector operation that does not involve any learnable parameters. For completeness, in a multi-head setting, where multiple target nodes $\bm{\nu} := \{v_k\}_{k=1}^K$ are considered, the output embedding can either be fed into a QNN and processed as a quantum analogue of a dense layer or again it can be combined classically through a simple concatenation, yielding 
\begin{align}
    \mathbf{h}^{(L)}_{\bm{\nu}}(\bm{\theta}) = \sigma\left(\left[\bigg\|_{k=1}^{K} \mathbf{h}_{v_k}^{(L)}(\bm{\theta}_v)\right]\right),
\end{align}
where $\bm{\theta} := \{ \bm{\theta}_{v_k} \}_{k=1}^{K}$. For simplicity, we assume target nodes have the same number of neighbors and that all QGCN circuits (or Ansätze) have the same depth $L$ across hops.

To this point, our analysis has focused on the unsupervised node-level task within the graph. By repeatedly performing message passing until the designated target node is reached, the resulting embedding captures both the intrinsic features of the target node and its topological relations within the graph. This final embedding yields the predicted graph-level output $\bar{y}(\bm{\theta}_v)$, where $\bm{\theta}_v$ denotes the trainable parameters of the QGCN circuit. This formulation then enables supervised learning by comparing the model’s prediction with the true target value $y$. The loss function $\mathcal{L}$ quantifies the training error, and the parameters $\bm{\theta}_v$ are optimized by minimizing $\mathcal{L}(\bar{y}(\bm{\theta}_v), y)$ as
\begin{align}
    \bm{\theta}^*_v &= \arg\min_{\bm{\theta}_v} \mathcal{L}(y(\bm{\theta}_v), y).
\end{align}
The parameters are optimized using classical gradient-based methods. Successful training yields a negligible prediction error, i.e. $|\bar{y}(\bm{\theta}_v) - y| \approx 0$.

\section{Results}\label{sec:results}
We now present the training and testing results of our QGNN framework on the QM9 dataset, comparing its performance with that of classical GNNs across molecules of varying sizes. Here, we use GNN to refer specifically to the GraphSAGE model.

\subsection{Implementation details}
Performance is evaluated using the Smooth L1 Loss and the R\textsuperscript{2} score, with an initial focus on a small subset of molecular samples. The selected Loss is chosen for its stability and robustness to outliers, making it particularly suitable for small sample sizes with high variability. Training employs the Adam optimizer~\cite{adam_2014} ($\beta_1 = \beta_2 = 0.9$) with a decaying learning rate.

Molecules are processed in two stages: (i) an unsupervised atom-level aggregation, where atomic features are combined according to the molecular adjacency matrix to construct a global embedding of the molecule, and (ii) a supervised molecule-level stage, where this embedding is used to predict a target property that characterizes the entire molecule. Here, the loss function compares this prediction with true targets, guiding the training of the model to capture both local and global structural information. Algorithm~\ref{alg:gnn} provides a high-level description of these steps. For the quantum models, the FM and the Ansatz are integrated directly into the model $\mathcal{M}$ definition.

{\footnotesize
\begin{algorithm}[H]
\caption{(Q)GNNs on the QM9 dataset}
\label{alg:gnn}
\begin{algorithmic}[1]
\Require Batch of molecules $G$ with adjacencies $A$, atom features $F$, targets $T$, model $\mathcal{M}$
\Ensure $T_\mathrm{true}$, $T_\mathrm{pred}$, $loss$
\State $loss \gets 0.0$, $T_\mathrm{pred} \gets [\,]$, $T_\mathrm{true} \gets [\,]$
\For{\textbf{each} mol $(A, F, T)$ in $\texttt{batch}$}
    \State $prev\_out \gets \mathbf{0}$, $atom\_outs \gets [\,]$
    \For{\textbf{each} node $v$ in $G$}
        \State $x \gets \texttt{concat}(F[A[v]], prev\_out)$
        \State $out \gets \mathcal{M}(x)$
        \State $prev\_out \gets out$, 
        \State Append $out$ to $atom\_outs$
    \EndFor
    \State $mol\_out \gets \texttt{mean}(atom\_outs)$
    \State Append $mol\_out$ to $T_\mathrm{pred}$
    \State Append $T$ to $T_\mathrm{true}$
    \State $loss \mathrel{+}= \texttt{criterion}(mol\_out, T) / |\texttt{batch}|$
\EndFor
\State \Return $T_\mathrm{true}$, $T_\mathrm{pred}$, $loss$
\end{algorithmic}
\end{algorithm}
}

In the proposed algorithm, all hops $k$ up to the number of atoms are considered; for instance, a 9-atom molecule undergoes 9 hops, with embeddings propagated from lower- to higher-index atoms, yielding a single-branch computational graph with one final head. Classical GNNs are implemented in \textsc{PyTorch}~\cite{pytorch}, while QGNNs use \textsc{Qadence}~\cite{qadence2025}. The number of inputs/qubits are 8, matching the 7 node feature dimension plus one (accounting for the mean self-embedding of the target node)\footnote{Features such as atomic number, chirality, degree, formal charge, radical electrons, hybridization, and scaled mass are kept, while aromaticity, hydrogen count, ring membership, and valence are excluded as they are consistently zero in the selected samples.}. QM9 data are encoded via the Fourier Feature Map. Measurements assess local magnetization with tunable parameters $\bm{\omega} := \{ \omega_{i} \}_{i=1}^{N}$, and the average total magnetization is defined as an expectation value 
\begin{align}
    \langle {\mathcal{O}}\rangle = \frac{1}{N}\sum_{i=1}^N \omega_i \langle Z^{(i)}\rangle.
    \label{eq:meas}
\end{align}

For the first set of results, and since our primary goal is to demonstrate that a single quantum circuit, namely a QGCN circuit such as the one shown in Fig.~\ref{fig:qgnn_circuit}, can handle graphs with varying numbers of nodes, we adopt a minimal quantum circuit design in which the QGCN serves as the sole aggregation mechanism. For each result, the same fixed-architecture circuit is applied at every aggregation step. Its structure (circuit depth) and number of input qubits (circuit width) remain constant within each implementation, even as the number of nodes varies across graphs, leading to a different number of nodes being input to the aggregator at each hop. Despite our chosen circuit design, the QGNN framework demonstrates reasonable training performance and, notably, enhanced generalization on unseen data relative to its classical counterpart.

\subsection{Comparison Between GNNs and QGNNs}
We performed numerical simulations to compare classical GNNs with our proposed QGNNs using two cases based on molecular size and sample count, evaluated by Loss and R² Score. These cases are:

\vspace{2mm}
\begin{itemize}[noitemsep,nolistsep]
    \item \textsc{Case 1:} a sample of 30 molecules with up to 9 atoms.
    \item \textsc{Case 2:} a sample of 30 molecules with up to 18 atoms.
\end{itemize}
\vspace{2mm}

\begin{figure}[H]
\centering
    \begin{subfigure}{0.45\textwidth}
        \includegraphics[width=\textwidth]{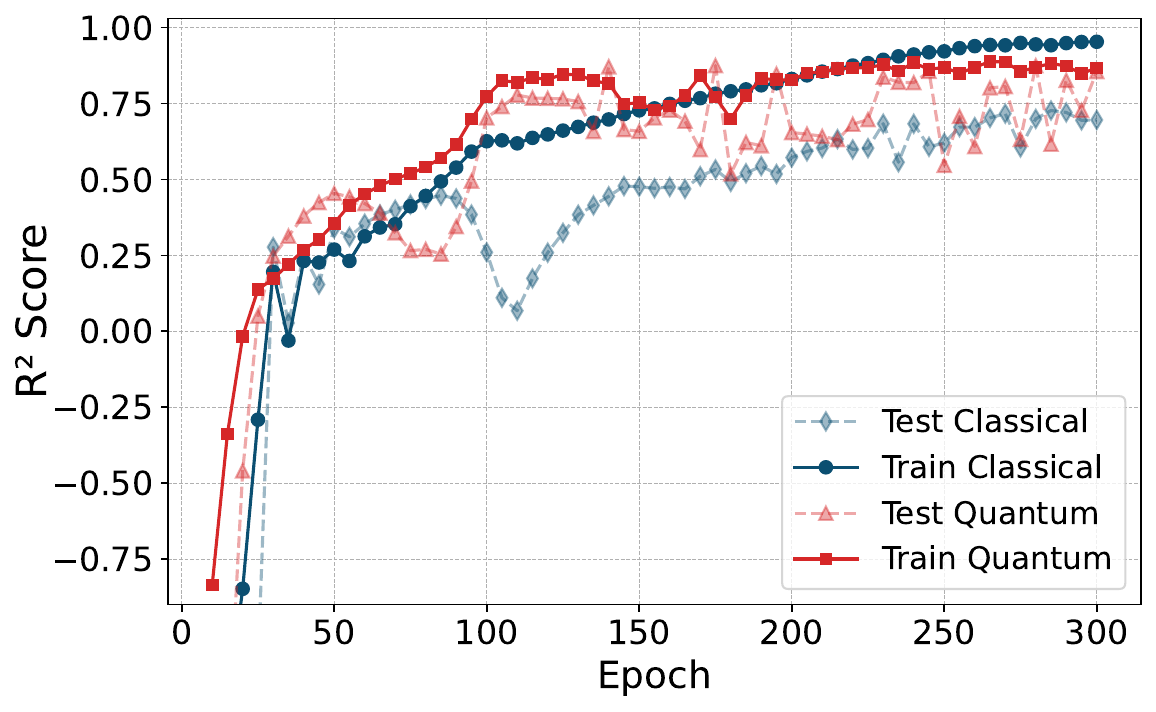}
        \vspace{-5mm}
        \caption{R² score curves  for Case 1}
        \vspace{2mm}
        \label{fig:7_score}
    \end{subfigure}
    \begin{subfigure}{0.45\textwidth}
        \includegraphics[width=\textwidth]{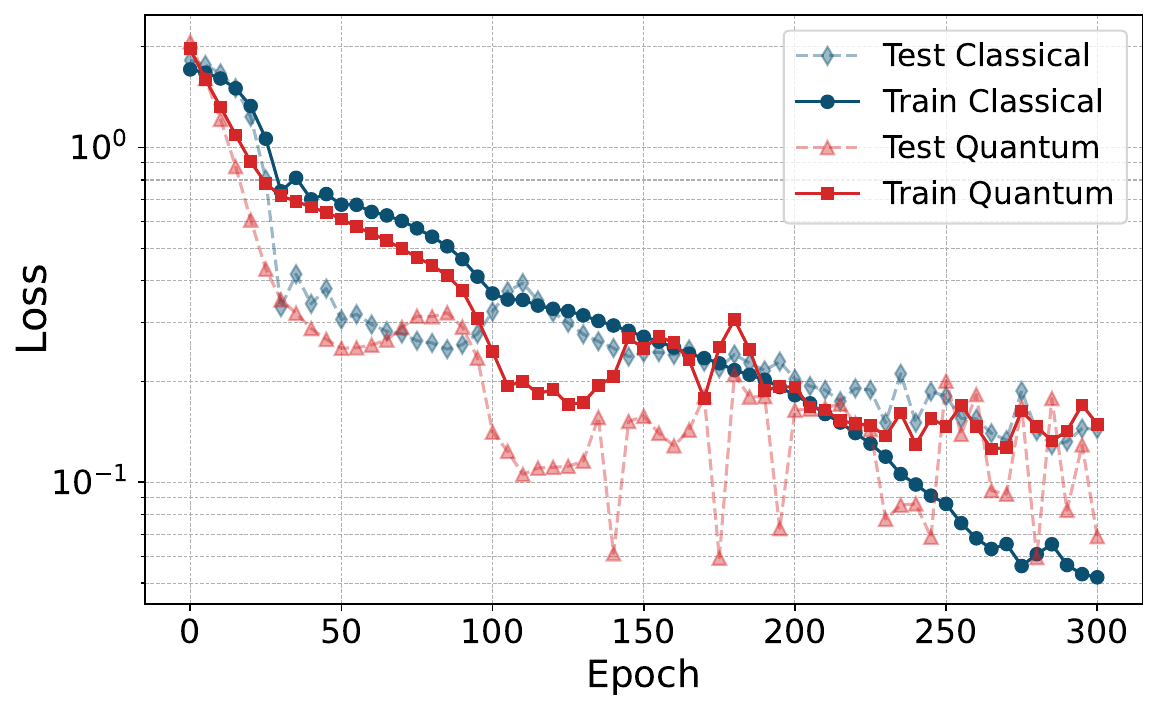}
        \vspace{-5mm}
        \caption{Loss curves for Case 1}
        \label{fig:7_loss}
    \end{subfigure}
    \caption{Comparison of training and testing performance for GNN (classical) and QGNN (quantum) models on molecules from the QM9 dataset, with all results obtained over 300 epochs. The markers are accompanied by dashed lines in their corresponding colors: circular and diamond markers distinguish the GNN's train and test results, respectively, while square and triangular markers represent the QGNN's train and test results. (a) and (b) present the R² score and Loss respectively for Case 1.}
    \label{fig:results_7_8_9}
\end{figure}

\begin{figure}[H]
    \centering
    \begin{subfigure}{0.45\textwidth}
        \includegraphics[width=\textwidth]{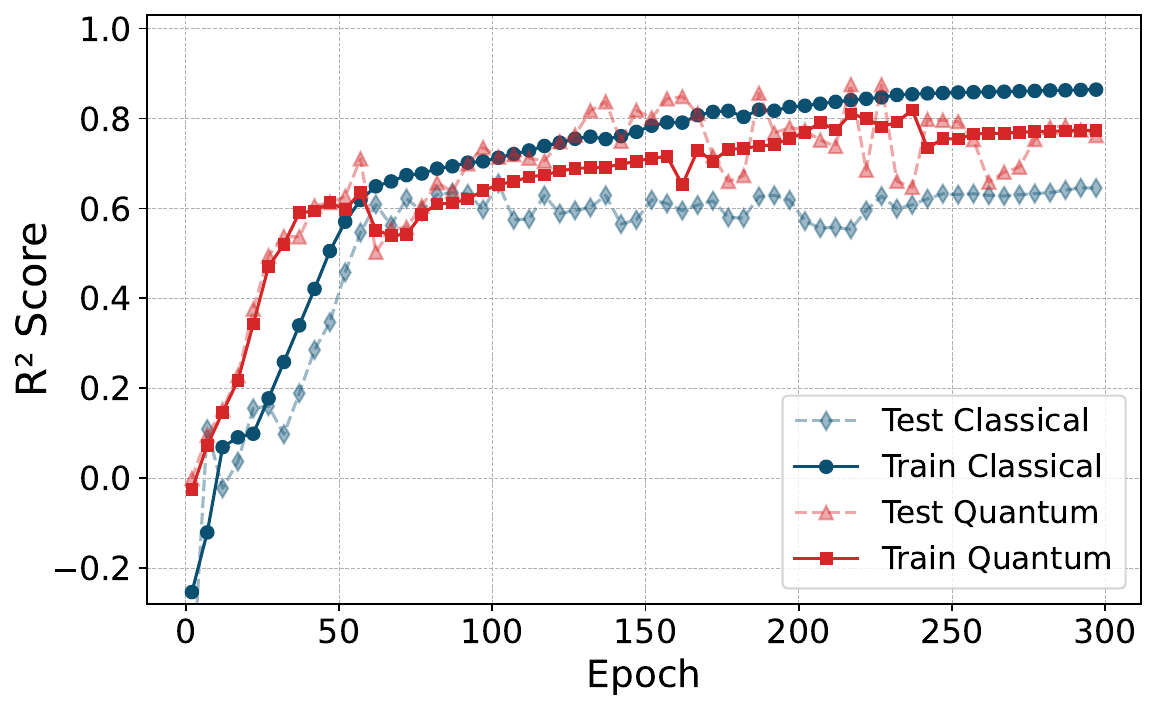}
         \vspace{-5mm}
         \caption{R² score curves  for Case 2}
         \vspace{2mm}
        \label{fig:8_score}
    \end{subfigure}
    \begin{subfigure}{0.45\textwidth}
        \includegraphics[width=\textwidth]{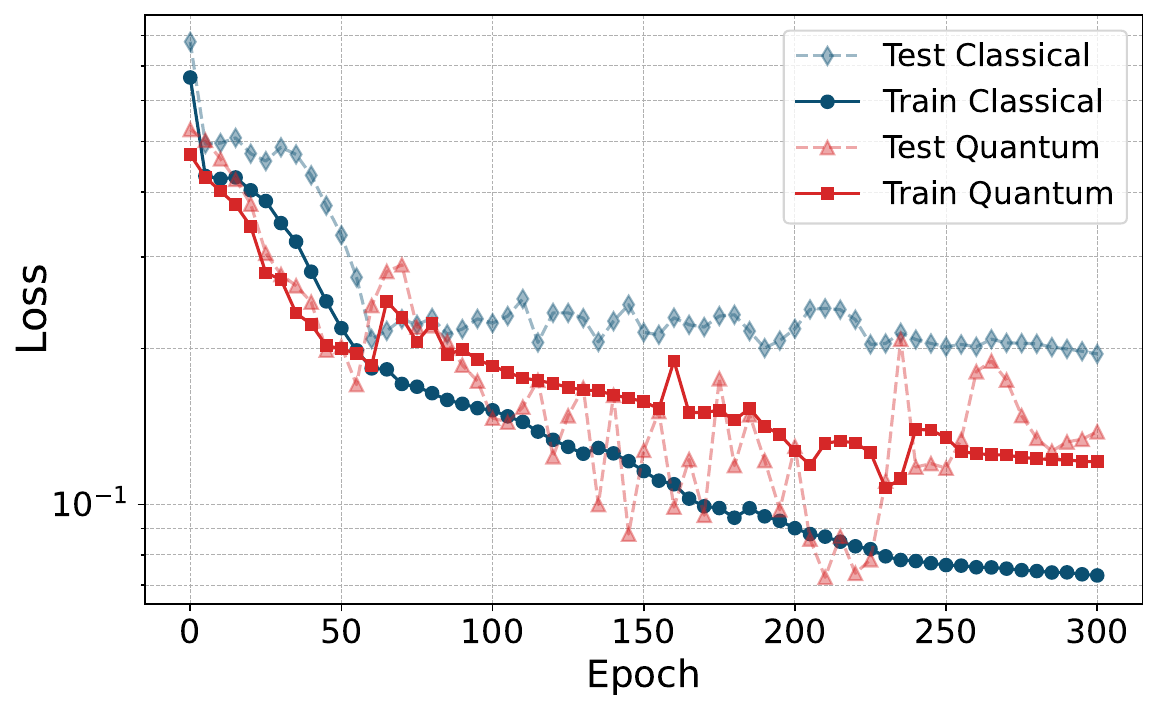}
        \vspace{-5mm}
        \caption{Loss curves for Case 2}
        \label{fig:8_loss}
    \end{subfigure}  
    \caption{Comparison of training and testing performance for GNN (classical) and QGNN (quantum) models on molecules from the QM9 dataset, with all results obtained over 300 epochs. The markers are accompanied by dashed lines in their corresponding colors: circular and diamond markers distinguish the GNN's train and test results, respectively, while square and triangular markers represent the QGNN's train and test results. (a) and (b) present the R² score and Loss respectively for Case 2.}
    \label{fig:results_7_8_91}
\end{figure}

\noindent Each case is sampled randomly and independently, with 70\% of the sample dataset used for training and 30\% for testing. The hyperparameters of the quantum and classical models used in the results are detailed in Table~\ref{tab:comparison}. Unless otherwise specified, the models hyperparameters and evaluation metrics: Smooth L1 Loss and $R^2$ score, remain unaltered. Furthermore, the classical model alternates convolutional layers with ReLU activations, followed by mean pooling and a final linear layer, mirroring the convolution-pooling structure of the QGCN so the comparison isolates the aggregation mechanism itself. It is also constructed to match the number of learnable parameters of the quantum model, ensuring a fair comparison between the two frameworks. All experiments are conducted over 300 epochs and the results are presented in Figs.~\ref{fig:results_7_8_9} and~\ref{fig:results_7_8_91}  for each case, showing the training and testing results of the GNN and QGNN frameworks. Table~\ref{tab:results} reports the best test-set results, which we use to assess generalization performance. The corresponding training loss and score are also included.

\begin{table}[H]
\centering
    \begin{tabular}{lccccc}
        \toprule
            \hspace{2mm}& FW & test R² & test Loss & train R² & train Loss \\ 
        \midrule
        \multirow{2}{*}{Case 1}\hspace{2mm}
            & GNN  & 0.7501 & 0.1178 & 0.9526 & 0.0530 \\
            & QGNN & 0.8912 & 0.0518 & 0.8525 & 0.1644 \\
        \midrule
        \multirow{2}{*}{Case 2} \hspace{2mm} 
            & GNN  & 0.6499 & 0.1951 & 0.8644 & 0.0730 \\
            & QGNN & 0.8801 & 0.0693 & 0.7917 & 0.1157 \\
        \bottomrule
    \end{tabular}
    \caption{Performance comparison of GNN and QGNN frameworks (FWs) across Cases 1 and 2. The table highlights the train and test R² scores, and losses for each model and case.}
    \label{tab:results}
\end{table}

\vspace{-1mm}
\begin{table}[H]
    \begin{tabular}{lccccccc}
        \toprule
         & samples & FW & inp/qub & \hspace{-2mm}FM & LR & $r$/layers & params \\
        \midrule
        \multirow{2}{*}{Case 1} 
            & 30 & GNN  & 8 \hspace{-5mm}& -   \hspace{1mm} & 0.01 & $[9,2]$ & 284 \\
            & 30 & QGNN   & 8 \hspace{-5mm}& Fourier \hspace{1mm} & 0.01& $[3,5]$ & 293 \\
        \midrule
        \multirow{2}{*}{Case 2}  \hspace{-1mm}
            \hspace{-5mm}& 30 & GNN  & 8 \hspace{-5mm}& -       \hspace{1mm} & 0.03 & $[8,4]$ & 305 \\
            & 30 & QGNN   & 8 \hspace{-5mm}& Fourier \hspace{1mm} & 0.03 & $[3,3,3]$ & 293 \\
        \bottomrule
    \end{tabular}
    \caption{The table reports the hyperparameters of the classical GNN and quantum QGNN frameworks (FWs) for Case~1 and Case~2. For each configuration, we list the number of training samples, the number of classical input features (inp) or qubits (qub), the feature map type (FM, quantum models only), the initial learning rate (LR), the hidden-layer architecture for classical models or the circuit depth configuration $r$ for quantum models, and the total number of trainable parameters (params).}
    \label{tab:comparison}
\end{table}

For Case~1, shown in Fig.~\ref{fig:results_7_8_9}, the classical GNN achieves superior training performance, whereas the QGNN exhibits improved generalization on the test set. In Case~2, Fig.~\ref{fig:results_7_8_91}, which includes molecules with up to twice as many atoms as in Case~1, the classical model again outperforms during training, while the quantum model maintains improved generalization. Notably, despite the substantial increase in molecular size between the two cases, the number of input qubits, and consequently the width of the quantum model, remains fixed. Moreover, within each case, the same quantum circuit is applied to all graphs, resulting in an unchanged circuit depth and consequently number of trainable parameters per case. This holds even though Case~2 exhibits greater variability in graph size, with molecules ranging from 7 to 18 atoms, suggesting a degree of robustness in handling larger graphs without requiring architectural modifications to the circuit.

Although the classical GNN consistently achieves lower training loss, the quantum model exhibits faster initial learning, particularly within the first 100 epochs in Case~1 and the first 50 epochs in Case~2 (see App.~\ref{app:seed_robustness}, where this effect is tested under different initialization). As molecular complexity increases from Case~1 to Case~2, the quantum model retains a strong ability to generalize to unseen data despite a decrease in training performance. This decrease, observed in both classical and quantum models, is likely attributable to the use of a single (Q)GCN for all aggregation steps when graphs of increased size variability are present in the dataset. Another possible justification for this decrease could be underfitting.

\noindent However, an underfitting model would be expected to perform poorly on both the training and test sets, whereas here the QGNN's test score remains high alongside its training score. We therefore interpret the lower training score not as a failure to learn, but as a consequence of the limited flexibility of the quantum circuit, which appears to act as an implicit regularizer. Notably, the classical GNN and QGNN are matched in parameter count (Table~ \ref{tab:comparison}) and thus have comparable raw expressivity, yet only the classical model shows this high-train/lower-test pattern. This asymmetry points to the architecture, rather than capacity, as the source of the difference, mirroring classical over-parameterized networks, where capacity measures based on parameter count fail to explain generalization behavior~\cite{zhang2017}, and implicit biases introduced by architecture and optimization play a central role instead~\cite{neyshabur2015}.

Despite the minimal design, these results highlight the potential of the proposed QGNN to process molecules of varying sizes and to offer advantages in generalization to unseen molecular structures. By benchmarking on a dataset as QM9,

\noindent this study motivates further investigation of  quantum convolutional  graph architectures.  Importantly, as is clearly shown in Figs.~\ref{fig:results_7_8_9} and ~\ref{fig:results_7_8_91}, the QGNN losses for both train and test set consistently plateaus during training across the two examined cases. Low model expressibility or the presence of suboptimal local minima in the loss landscape may be contributing factors that hinder the trainability of the quantum model \cite{anschuetz_2021_critical,bittel_2021_training}. To rule out the possibility that the QGNN's performance advantage stems from the Fourier feature encoding rather than the variational quantum circuit itself, we conducted a study with classical GNN baselines augmented with the same encoding, confirming that the encoding alone does not account for the observed gap (see App.~\ref{app:fourier_ablation}).
In addition, since the number of qubits is fixed by the feature dimension and each node aggregates a bounded number of neighbors under our GraphSAGE-style sampling scheme, the classical simulation cost per node remains constant. The QCNN-inspired ansatz's logarithmic depth~\cite{pesah_physrevX_2021} keeps this constant overhead shallow, and the overall cost scales linearly with graph size, matching classical GNNs. We emphasize that the contribution of this work lies in the use of genuinely quantum Ans\"{a}tze to generate node embeddings across graphs of varying size, and in the enhanced generalization this enables, rather than in any runtime or complexity advantage.

\subsection{Multiple quantum aggregators}
\begin{figure}[H]
    \begin{subfigure}{0.48\textwidth}
        \includegraphics[width=\textwidth]{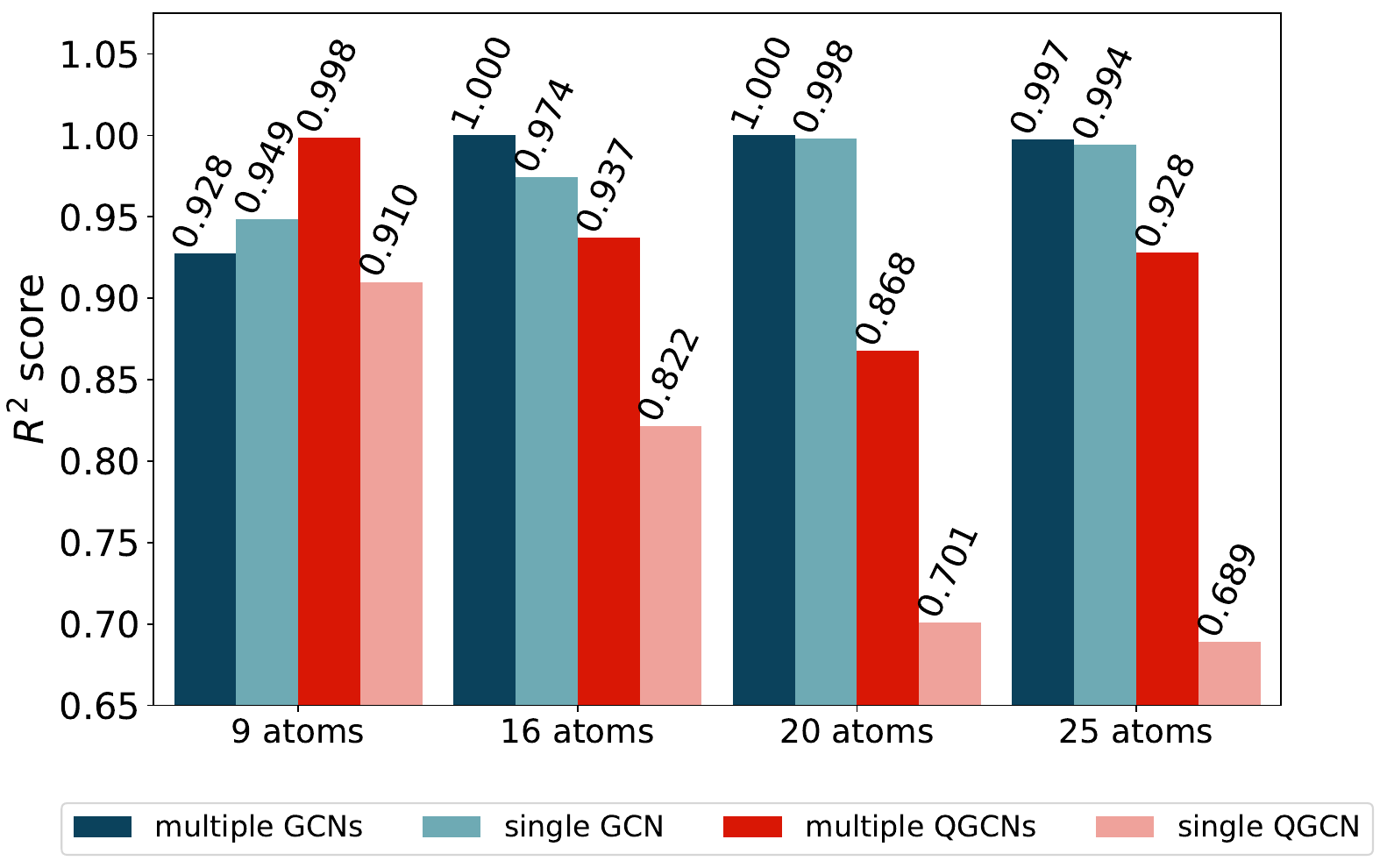}
        \caption{R² scores}
        \vspace{2mm}
    \end{subfigure}
    \begin{subfigure}{0.48\textwidth}
        \includegraphics[width=\textwidth]{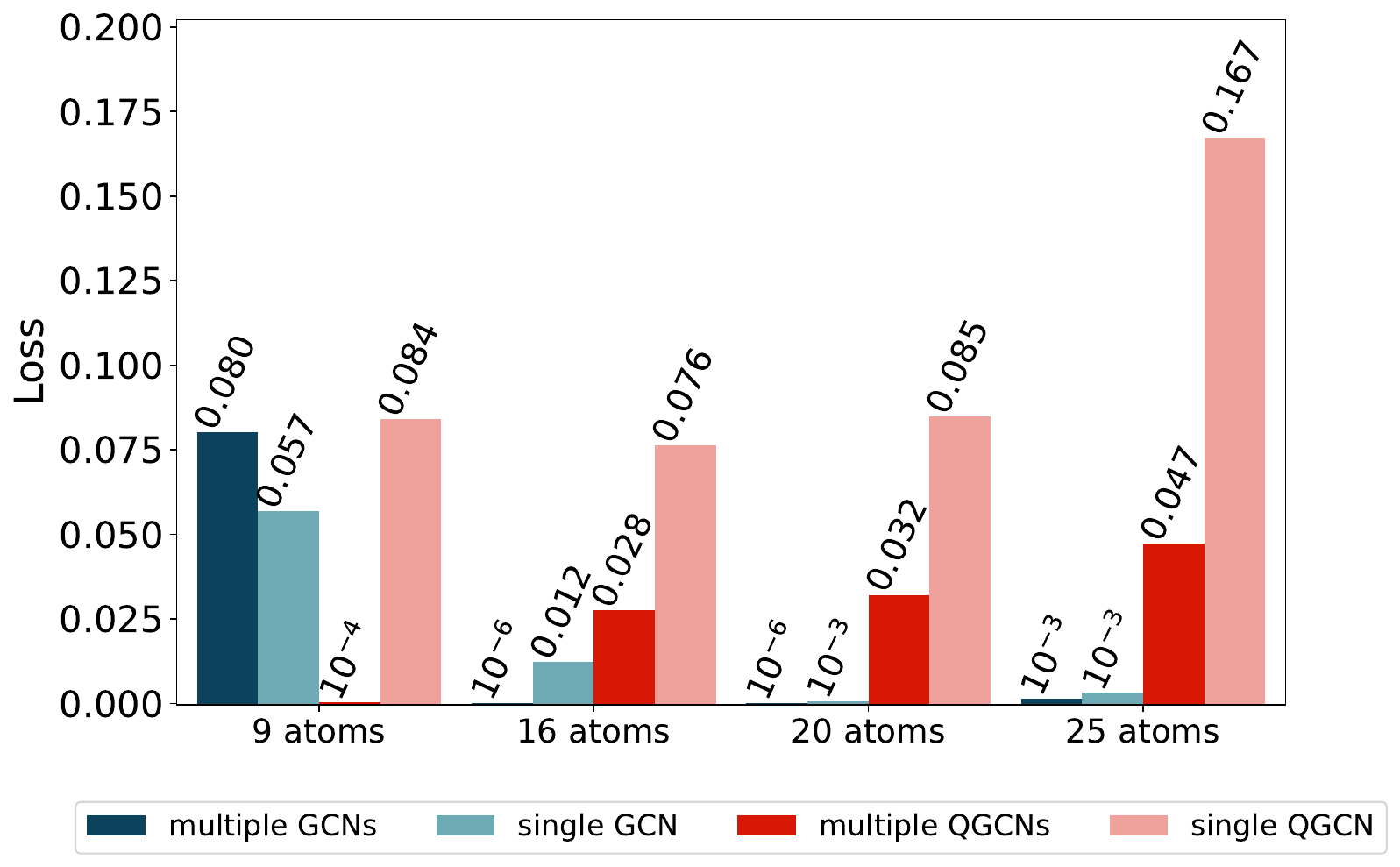}
        \caption{Losses}
    \end{subfigure}
    \caption{Training R$^2$ scores (a) and losses (b) for multiple GCNs, single GCN, multiple QGCNs, and single QGCN across molecular sets ranging from $\leq 9$ to $\leq 25$ atoms, trained for 300 epochs. Parameter counts are reported in Table~\ref{tab:params_ablation}.}
    \label{fig:multi}
\end{figure}

To address the observed limitations in trainability, we investigate the use of multiple quantum aggregators (QGCN circuits) in our QGNN framework, using molecules with up to 9, 16, 20, and 25 atoms (30 independent random samples per case). Instead of reusing a single shared quantum circuit as the aggregator (see the schematic representation in~\ref{fig:comp_graph}), we assign a distinct, shallow quantum circuit to each hop, so that the number of circuits matches the maximum hop depth for a given sample; for graphs with up to 9 atoms, for example, this yields at most 9 distinct circuits, each a single convolution-pooling layer ($r=[1]$). This design increases the total number of tunable parameters and model expressibility relative to a single shared circuit, while the shallow per-hop circuits remain individually easy to optimize, improving both training stability and efficiency. Since aggregators are assigned by hop depth rather than by node identity, permutation equivariance is preserved: relabeling the graph's nodes or edges leaves the hop structure, and hence the prediction, unchanged, consistent with classical frameworks such as GraphSAGE~\cite{leskovec_inductive_representation_2018} and GATs~\cite{petar_gats_2018}. Figure~\ref{fig:multi} shows the results, obtained with an initial learning rate of $0.03$, and confirms that the QGNN benefits substantially from this multi-QGCN design: training loss decreases and training score increases across all cases, including molecules of up to 25 atoms. For a fair comparison, we also train a single QGCN whose parameter count is matched to that of the corresponding multiple-QGCN model in each case (Table~ \ref{tab:params_ablation}), using convolution-pooling configurations $r=[7,7,5]$, $r=[13,11,9]$, $r=[15,15,13]$, and $r=[19,19,17]$, respectively.

For completeness, we repeat this comparison classically, using multiple GCNs and a single GCN baseline. The multiple GCNs use hidden layers $[2,2]$ per aggregator across all cases; the single GCN uses hidden layers $[8,8,8]$, $[12,11,11]$, $[13,13,12]$, and $[19,13,11]$ for molecules with up to 9, 16, 20, and 25 atoms, respectively, with all models trained at the same initial learning rate of $0.03$. Unlike the quantum case, the multiple GCNs bring no improvement over the single GCN baseline, and even degrade performance for molecules with up to 9 atoms; in this same case, the multiple QGCNs achieve the best results of all four configurations, in both R$^2$ score and loss. This suggests that the benefit of multiple aggregators is not a general architectural property, but specific to the quantum setting, where each hop generates local quantum-native node embeddings that appear to capture richer structural context than their classical counterparts in this regime.

\begin{table}[H]
        \centering
        \begin{tabular}{lcccc}
            \hline
            \textbf{Model} & \textbf{9 atoms} & \textbf{16 atoms} & 
            \textbf{20 atoms} & \textbf{25 atoms} \\
            \hline
            Multiple GCNs  & 602 & 1072 & 1340 & 1675 \\
            Single GCN     & 609 & 1093 & 1338 & 1681\\
            Multiple QGCNs & 612 & 1088 & 1360 & 1700 \\
            Single QGCN    & 623 & 1088 & 1343 & 1703 \\
            \hline
                \end{tabular}
        \caption{Trainable parameters for the Multiple QGCN and parameter-matched Single QGCN across molecular sizes.}
        \label{tab:params_ablation}
    \end{table}

An alternative approach to mitigate the observed loss stagnation and trainability issues observed in Figs.~\ref{fig:results_7_8_9}–\ref{fig:results_7_8_91} is to increase the number of qubits used to encode the classical node features, thereby enhancing the expressibility of the quantum model. However, this strategy may introduce additional challenges, most notably the emergence of barren plateaus, a well-documented issue in quantum machine learning that can further hinder trainability~\cite{cerezo_nature_2021_costfunc, larocca_2024_review}. Nevertheless, recent studies on QCNNs indicate that barren plateaus can be avoided under certain architectural conditions~\cite{pesah_physrevX_2021, cerezo_provable_absence_2024}. Since the QGCN circuitry is inspired by QCNNs, adopting similar design principles may advance QGNNs toward a more reliable and scalable framework. It is worth noting that although the absence of BPs in QCNNs was established previously~\cite{pesah_physrevX_2021}, it is not guaranteed that this property extends to graph-structured, permutation-equivariant settings; we verify this numerically in the following section for our framework.

Next, we investigate whether the previously discussed strategies can effectively mitigate or prevent barren plateaus in our proposed QGNN framework.

%

\subsection{Scalability and absence of barren plateaus in QGNNs}
Barren plateaus~\cite{anschuetz_2022_beyond, cerezo_2020_impact, larocca_2024_review, mcclean_2018_barren}
pose a significant challenge to the trainability of QML models~\cite{ragone_2023_lie, holmes_2021_connecting, bermejo_2024_improving}, as they lead to an exponential decay in the gradients of the loss function. The absence of barren plateaus implies that minimizing the loss function landscape does not require exponentially high precision, allowing expectation values to be measured with polynomial precision relative to the system size. Here, we numerically confirm the scalability of the proposed QGNN framework, as barren plateaus are absent with increasing number of qubits. This ensures continued trainability as  the model scales. To assess the presence of barren plateaus, we evaluate the variance of the loss function gradient, $\operatorname{Var}[\partial_{\theta_\mu}\mathcal{L}]$, where $\theta_\mu$ denotes a variational parameter, by sampling multiple points across the loss landscape. According to Ref.~\cite{arrasmith_2021_equivalence}, we have the following Chebyshev's inequality
\begin{align}
    P\left(\left|\partial_{\theta_{\mu}} \mathcal{L}\right| \geq \delta\right) \leq \frac{\operatorname{Var}[\partial_{\theta_\mu} \mathcal{L}]}{\delta^2}.
    \label{eq:cheb_bound}
\end{align}
where $\delta>0$ is a positive threshold. Thus, if the variance of the loss function partial derivative decreases exponentially, the probability that the partial derivative is non-zero similarly vanishes, indicating a barren plateau. 

Inspired by Ref.~\cite{pesah_physrevX_2021}, we study the gradient scaling of our QGCN architecture under the same conditions. Specifically, as shown in the schematic representation of the circuit.~\ref{fig:ansatz}, we consider two convolutional layer designs: an uncorrelated construction composed of independent local 2-designs $W$, and a correlated variant in which all $W$ operators within a layer share the same parameters. Finally, the loss function is defined by local measurements, ensuring its local nature due to the pooling structure of quantum convolutional models.

\begin{figure}[H]
    \includegraphics[width=0.45\textwidth]{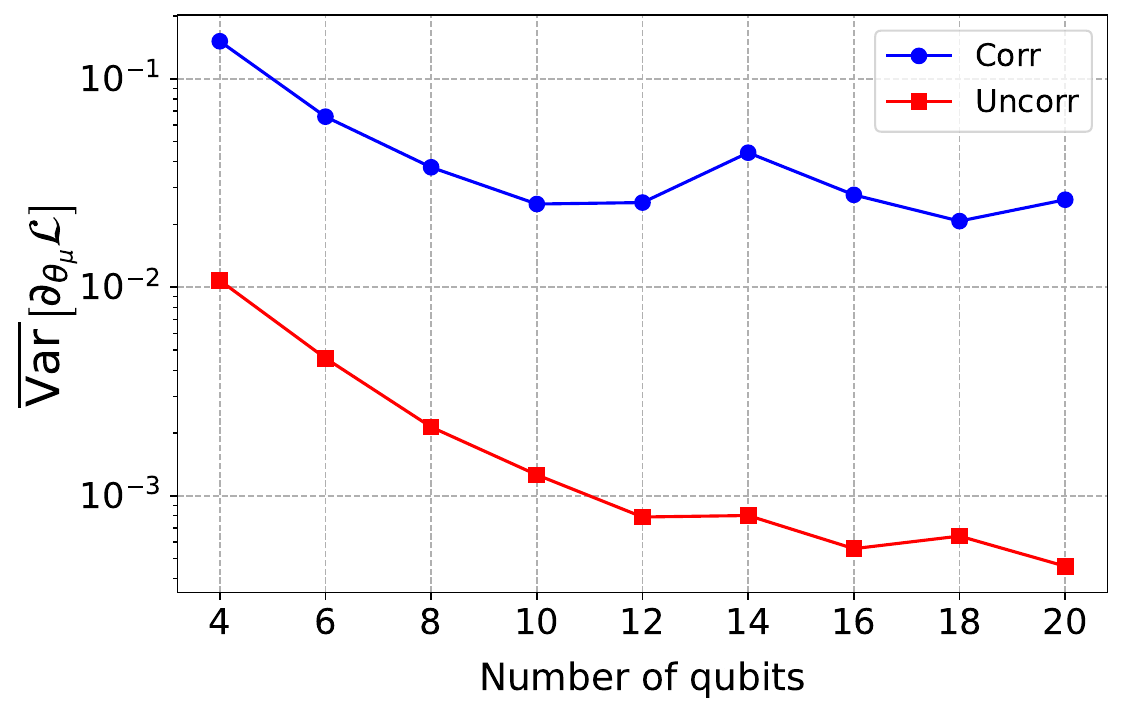}
    \caption{Average variance of the loss function partial derivatives $\overline{\operatorname{Var}}\left[\partial_{\theta_\mu} \mathcal{L}\right]$ of the QGCN circuit. The correlated QGNN (corr) uses the same parameters for all $W$ cells in a convolution layer, while in the uncorrelated (uncorr) all cells are independent. The variance for each variational parameter $\theta_\mu$ is calculated over 200 points in the loss landscape.}
    \label{fig:var_grads}
\end{figure}

The numerical results for the variance of the loss gradients are shown in Fig.~\ref{fig:var_grads}. The plot displays the average variance $\overline{\operatorname{Var}}\left[\partial_{\theta_\mu} \mathcal{L}\right] = \sum^{N}_{\mu}\operatorname{Var}\left[\partial_{\theta_\mu} \mathcal{L}\right]/N$ across all $N$ variational parameters for each QGCN architecture (correlated and uncorrelated). Unlike previous results, we use synthetic data as input node features to allow us to easily increase their dimensionality and, consequently, the number of qubits. By sampling 200 random points on the loss landscape, we compute the average variance for each parameter $\theta_\mu$, treating each number of qubit case separately. Remarkably, both the correlated and uncorrelated Ansätze suggest sub-exponential scaling of the variance, confirming the absence of barren plateaus in both models. In particular, the correlated Ansatz achieves higher variances that remain nearly constant as the number of qubits increases. This suggests that introducing correlation in the parameters can enhance the trainability of the QGNN model at scale.  Importantly, in this particular analysis, we employ a single QGCN circuit, so the absence of barren plateaus in the circuit also implies its absence in the QGNN framework. The absence of barren plateaus in this QGNN framework establishes a trainable baseline for exploring more robust and scalable use cases. Here we conclude our analysis.

\section{Conclusions}\label{sec:conclusions}
We presented a novel Quantum Graph Neural Network framework that generates node embeddings by combining inductive representation learning with quantum circuit design, providing a flexible architecture that lays the foundation for improved generalization across graphs of varying size and structure. Specifically, our QGNN integrates sampling and aggregation techniques from GraphSAGE, using QGCNs as aggregators. The QGCN circuit is inspired by previous QCNN designs~\cite{lukin_qcnn_2019}, featuring convolutional and pooling layers, while maintaining high flexibility to accommodate diverse architectures. By defining the convolutional cell, the circuit is uniquely determined, enabling the construction of tailored quantum circuits for specific problems or symmetry-preserving architectures.

Benchmarking on the QM9 dataset shows that while the classical GNN achieves superior training performance, the QGNN exhibits enhanced generalization, particularly as molecular complexity increases. In this evaluation, we employ a single, minimal QGCN circuit reused at every aggregation step. Hence, the quantum circuit architecture remains unchanged within each case, with tunable parameters staying constant regardless of the number of nodes in a graph, showcasing the ability to handle graphs with varying number of nodes without requiring design modifications in the circuit. Additionally, the quantum approach shows faster initial learning, indicating potential for rapid convergence. To further address trainability limitations observed in the single-circuit setup, we also investigate an extension of the proposed QGNN framework in which each aggregation step employs a distinct quantum circuit. In this configuration, different QGCN circuits are assigned to each hop, which remarkably boosts training performance across all tested cases.

Finally, by investigating the scalability of the QGNN, we numerically confirm the absence of barren plateaus in the proposed framework as the number of qubits increases. This suggests its potential for further improvement to tackle larger and more complex graph-based problems. More broadly, this work establishes the framework in a noise-free setting as a baseline; evaluating its performance under noisy simulations, and eventually on NISQ hardware, is left for future work.

\vspace{-3mm}

\subsection*{Acknowledgments}
\vspace{-4mm}

The authors thank Mehdi Djellabi, Sachin Kasture, Igor O. Sokolov, Shaheen Acheche, and Antonio Andrea Gentile for very useful discussions and comments on early drafts. AMF. acknowledges the project ``Divide \& Quantum'' (P.N. 1389.20.241) of the NWA-ORC programme (partly) financed by the Dutch Research Council (NWO). IFG \& SV acknowledge funding via ``Qu\&Co flow'' project ID (EC): 190146292. 

\bibliography{qgnn} 

\begin{thebibliography}{65}%
\makeatletter
\providecommand \@ifxundefined [1]{%
 \@ifx{#1\undefined}
}%
\providecommand \@ifnum [1]{%
 \ifnum #1\expandafter \@firstoftwo
 \else \expandafter \@secondoftwo
 \fi
}%
\providecommand \@ifx [1]{%
 \ifx #1\expandafter \@firstoftwo
 \else \expandafter \@secondoftwo
 \fi
}%
\providecommand \natexlab [1]{#1}%
\providecommand \enquote  [1]{``#1''}%
\providecommand \bibnamefont  [1]{#1}%
\providecommand \bibfnamefont [1]{#1}%
\providecommand \citenamefont [1]{#1}%
\providecommand \href@noop [0]{\@secondoftwo}%
\providecommand \href [0]{\begingroup \@sanitize@url \@href}%
\providecommand \@href[1]{\@@startlink{#1}\@@href}%
\providecommand \@@href[1]{\endgroup#1\@@endlink}%
\providecommand \@sanitize@url [0]{\catcode `\\12\catcode `\$12\catcode `\&12\catcode `\#12\catcode `\^12\catcode `\_12\catcode `\%12\relax}%
\providecommand \@@startlink[1]{}%
\providecommand \@@endlink[0]{}%
\providecommand \url  [0]{\begingroup\@sanitize@url \@url }%
\providecommand \@url [1]{\endgroup\@href {#1}{\urlprefix }}%
\providecommand \urlprefix  [0]{URL }%
\providecommand \Eprint [0]{\href }%
\providecommand \doibase [0]{https://doi.org/}%
\providecommand \selectlanguage [0]{\@gobble}%
\providecommand \bibinfo  [0]{\@secondoftwo}%
\providecommand \bibfield  [0]{\@secondoftwo}%
\providecommand \translation [1]{[#1]}%
\providecommand \BibitemOpen [0]{}%
\providecommand \bibitemStop [0]{}%
\providecommand \bibitemNoStop [0]{.\EOS\space}%
\providecommand \EOS [0]{\spacefactor3000\relax}%
\providecommand \BibitemShut  [1]{\csname bibitem#1\endcsname}%
\let\auto@bib@innerbib\@empty
\bibitem [{\citenamefont {Borisyuk}\ \emph {et~al.}(2024)\citenamefont {Borisyuk}, \citenamefont {He}, \citenamefont {Ouyang}, \citenamefont {Ramezani}, \citenamefont {Du}, \citenamefont {Hou}, \citenamefont {Jiang}, \citenamefont {Pasumarthy}, \citenamefont {Bannur}, \citenamefont {Tiwana} \emph {et~al.}}]{borisyuk_2024_lignn}%
  \BibitemOpen
  \bibfield  {author} {\bibinfo {author} {\bibfnamefont {F.}~\bibnamefont {Borisyuk}}, \bibinfo {author} {\bibfnamefont {S.}~\bibnamefont {He}}, \bibinfo {author} {\bibfnamefont {Y.}~\bibnamefont {Ouyang}}, \bibinfo {author} {\bibfnamefont {M.}~\bibnamefont {Ramezani}}, \bibinfo {author} {\bibfnamefont {P.}~\bibnamefont {Du}}, \bibinfo {author} {\bibfnamefont {X.}~\bibnamefont {Hou}}, \bibinfo {author} {\bibfnamefont {C.}~\bibnamefont {Jiang}}, \bibinfo {author} {\bibfnamefont {N.}~\bibnamefont {Pasumarthy}}, \bibinfo {author} {\bibfnamefont {P.}~\bibnamefont {Bannur}}, \bibinfo {author} {\bibfnamefont {B.}~\bibnamefont {Tiwana}}, \emph {et~al.},\ }\bibfield  {title} {\bibinfo {title} {Lignn: Graph neural networks at linkedin},\ }\href {https://arxiv.org/abs/2402.11139} {\bibfield  {journal} {\bibinfo  {journal} {arXiv:2402.11139 [cs.LG]}\ } (\bibinfo {year} {2024})}\BibitemShut {NoStop}%
\bibitem [{\citenamefont {Fan}\ \emph {et~al.}(2019)\citenamefont {Fan}, \citenamefont {Ma}, \citenamefont {Li}, \citenamefont {He}, \citenamefont {Zhao}, \citenamefont {Tang},\ and\ \citenamefont {Yin}}]{fan_2019_graph}%
  \BibitemOpen
  \bibfield  {author} {\bibinfo {author} {\bibfnamefont {W.}~\bibnamefont {Fan}}, \bibinfo {author} {\bibfnamefont {Y.}~\bibnamefont {Ma}}, \bibinfo {author} {\bibfnamefont {Q.}~\bibnamefont {Li}}, \bibinfo {author} {\bibfnamefont {Y.}~\bibnamefont {He}}, \bibinfo {author} {\bibfnamefont {E.}~\bibnamefont {Zhao}}, \bibinfo {author} {\bibfnamefont {J.}~\bibnamefont {Tang}},\ and\ \bibinfo {author} {\bibfnamefont {D.}~\bibnamefont {Yin}},\ }\bibfield  {title} {\bibinfo {title} {Graph neural networks for social recommendation},\ }in\ \href@noop {} {\emph {\bibinfo {booktitle} {The world wide web conference}}}\ (\bibinfo {year} {2019})\ pp.\ \bibinfo {pages} {417--426}\BibitemShut {NoStop}%
\bibitem [{\citenamefont {De~Nadai}\ \emph {et~al.}(2024)\citenamefont {De~Nadai}, \citenamefont {Fabbri}, \citenamefont {Gigioli}, \citenamefont {Wang}, \citenamefont {Li}, \citenamefont {Silvestri}, \citenamefont {Kim}, \citenamefont {Lin}, \citenamefont {Radosavljevic}, \citenamefont {Ghael}, \citenamefont {Nyhan}, \citenamefont {Bouchard}, \citenamefont {Lalmas-Roelleke},\ and\ \citenamefont {Damianou}}]{de_2024_personalized}%
  \BibitemOpen
  \bibfield  {author} {\bibinfo {author} {\bibfnamefont {M.}~\bibnamefont {De~Nadai}}, \bibinfo {author} {\bibfnamefont {F.}~\bibnamefont {Fabbri}}, \bibinfo {author} {\bibfnamefont {P.}~\bibnamefont {Gigioli}}, \bibinfo {author} {\bibfnamefont {A.}~\bibnamefont {Wang}}, \bibinfo {author} {\bibfnamefont {A.}~\bibnamefont {Li}}, \bibinfo {author} {\bibfnamefont {F.}~\bibnamefont {Silvestri}}, \bibinfo {author} {\bibfnamefont {L.}~\bibnamefont {Kim}}, \bibinfo {author} {\bibfnamefont {S.}~\bibnamefont {Lin}}, \bibinfo {author} {\bibfnamefont {V.}~\bibnamefont {Radosavljevic}}, \bibinfo {author} {\bibfnamefont {S.}~\bibnamefont {Ghael}}, \bibinfo {author} {\bibfnamefont {D.}~\bibnamefont {Nyhan}}, \bibinfo {author} {\bibfnamefont {H.}~\bibnamefont {Bouchard}}, \bibinfo {author} {\bibfnamefont {M.}~\bibnamefont {Lalmas-Roelleke}},\ and\ \bibinfo {author} {\bibfnamefont {A.}~\bibnamefont {Damianou}},\ }\bibfield  {title} {\bibinfo {title} {Personalized audiobook recommendations at spotify through graph neural
  networks},\ }in\ \href {https://doi.org/10.1145/3589335.3648339} {\emph {\bibinfo {booktitle} {Companion Proceedings of the ACM Web Conference 2024 (WWW '24 Companion)}}}\ (\bibinfo {year} {2024})\ pp.\ \bibinfo {pages} {403--412}\BibitemShut {NoStop}%
\bibitem [{\citenamefont {Jain}\ \emph {et~al.}(2019)\citenamefont {Jain}, \citenamefont {Liu}, \citenamefont {Sarda},\ and\ \citenamefont {Molino}}]{jain_2019_food}%
  \BibitemOpen
  \bibfield  {author} {\bibinfo {author} {\bibfnamefont {A.}~\bibnamefont {Jain}}, \bibinfo {author} {\bibfnamefont {I.}~\bibnamefont {Liu}}, \bibinfo {author} {\bibfnamefont {A.}~\bibnamefont {Sarda}},\ and\ \bibinfo {author} {\bibfnamefont {P.}~\bibnamefont {Molino}},\ }\bibfield  {title} {\bibinfo {title} {Food discovery with uber eats: Using graph learning to power recommendations},\ }\href@noop {} {\bibfield  {journal} {\bibinfo  {journal} {Accessed March}\ }\textbf {\bibinfo {volume} {1}},\ \bibinfo {pages} {2021} (\bibinfo {year} {2019})}\BibitemShut {NoStop}%
\bibitem [{\citenamefont {Stokes}\ \emph {et~al.}(2020)\citenamefont {Stokes}, \citenamefont {Yang}, \citenamefont {Swanson}, \citenamefont {Jin}, \citenamefont {Cubillos-Ruiz}, \citenamefont {Donghia}, \citenamefont {MacNair}, \citenamefont {French}, \citenamefont {Carfrae}, \citenamefont {Bloom-Ackermann} \emph {et~al.}}]{stokes_2020_deep}%
  \BibitemOpen
  \bibfield  {author} {\bibinfo {author} {\bibfnamefont {J.~M.}\ \bibnamefont {Stokes}}, \bibinfo {author} {\bibfnamefont {K.}~\bibnamefont {Yang}}, \bibinfo {author} {\bibfnamefont {K.}~\bibnamefont {Swanson}}, \bibinfo {author} {\bibfnamefont {W.}~\bibnamefont {Jin}}, \bibinfo {author} {\bibfnamefont {A.}~\bibnamefont {Cubillos-Ruiz}}, \bibinfo {author} {\bibfnamefont {N.~M.}\ \bibnamefont {Donghia}}, \bibinfo {author} {\bibfnamefont {C.~R.}\ \bibnamefont {MacNair}}, \bibinfo {author} {\bibfnamefont {S.}~\bibnamefont {French}}, \bibinfo {author} {\bibfnamefont {L.~A.}\ \bibnamefont {Carfrae}}, \bibinfo {author} {\bibfnamefont {Z.}~\bibnamefont {Bloom-Ackermann}}, \emph {et~al.},\ }\bibfield  {title} {\bibinfo {title} {A deep learning approach to antibiotic discovery},\ }\href@noop {} {\bibfield  {journal} {\bibinfo  {journal} {Cell}\ }\textbf {\bibinfo {volume} {180}},\ \bibinfo {pages} {688} (\bibinfo {year} {2020})}\BibitemShut {NoStop}%
\bibitem [{\citenamefont {Derrow-Pinion}\ \emph {et~al.}(2021)\citenamefont {Derrow-Pinion}, \citenamefont {She}, \citenamefont {Wong}, \citenamefont {Lange}, \citenamefont {Hester}, \citenamefont {Perez}, \citenamefont {Nunkesser}, \citenamefont {Lee}, \citenamefont {Guo}, \citenamefont {Wiltshire} \emph {et~al.}}]{derrow_2021_eta}%
  \BibitemOpen
  \bibfield  {author} {\bibinfo {author} {\bibfnamefont {A.}~\bibnamefont {Derrow-Pinion}}, \bibinfo {author} {\bibfnamefont {J.}~\bibnamefont {She}}, \bibinfo {author} {\bibfnamefont {D.}~\bibnamefont {Wong}}, \bibinfo {author} {\bibfnamefont {O.}~\bibnamefont {Lange}}, \bibinfo {author} {\bibfnamefont {T.}~\bibnamefont {Hester}}, \bibinfo {author} {\bibfnamefont {L.}~\bibnamefont {Perez}}, \bibinfo {author} {\bibfnamefont {M.}~\bibnamefont {Nunkesser}}, \bibinfo {author} {\bibfnamefont {S.}~\bibnamefont {Lee}}, \bibinfo {author} {\bibfnamefont {X.}~\bibnamefont {Guo}}, \bibinfo {author} {\bibfnamefont {B.}~\bibnamefont {Wiltshire}}, \emph {et~al.},\ }\bibfield  {title} {\bibinfo {title} {Eta prediction with graph neural networks in google maps},\ }in\ \href@noop {} {\emph {\bibinfo {booktitle} {Proceedings of the 30th ACM International Conference on Information \& Knowledge Management}}}\ (\bibinfo {year} {2021})\ pp.\ \bibinfo {pages} {3767--3776}\BibitemShut {NoStop}%
\bibitem [{\citenamefont {Ngo}\ \emph {et~al.}(2023)\citenamefont {Ngo}, \citenamefont {Hy},\ and\ \citenamefont {Kondor}}]{kondor_modeling_polypharmacy_2023}%
  \BibitemOpen
  \bibfield  {author} {\bibinfo {author} {\bibfnamefont {N.~K.}\ \bibnamefont {Ngo}}, \bibinfo {author} {\bibfnamefont {T.~S.}\ \bibnamefont {Hy}},\ and\ \bibinfo {author} {\bibfnamefont {R.}~\bibnamefont {Kondor}},\ }\bibfield  {title} {\bibinfo {title} {Modeling polypharmacy and predicting drug-drug interactions using deep generative models on multimodal graphs},\ }\href {https://arxiv.org/abs/2302.08680} {\bibfield  {journal} {\bibinfo  {journal} {arXiv:2302.08680v1 [cs.LG]}\ } (\bibinfo {year} {2023})}\BibitemShut {NoStop}%
\bibitem [{\citenamefont {Zitnik}\ \emph {et~al.}(2018)\citenamefont {Zitnik}, \citenamefont {Agrawal},\ and\ \citenamefont {Leskovec}}]{zitnik_2018_modeling}%
  \BibitemOpen
  \bibfield  {author} {\bibinfo {author} {\bibfnamefont {M.}~\bibnamefont {Zitnik}}, \bibinfo {author} {\bibfnamefont {M.}~\bibnamefont {Agrawal}},\ and\ \bibinfo {author} {\bibfnamefont {J.}~\bibnamefont {Leskovec}},\ }\bibfield  {title} {\bibinfo {title} {Modeling polypharmacy side effects with graph convolutional networks},\ }\href {https://doi.org/10.1093/bioinformatics/bty294} {\bibfield  {journal} {\bibinfo  {journal} {Bioinformatics}\ }\textbf {\bibinfo {volume} {34}},\ \bibinfo {pages} {i457} (\bibinfo {year} {2018})}\BibitemShut {NoStop}%
\bibitem [{\citenamefont {Arute}\ \emph {et~al.}(2019)\citenamefont {Arute}, \citenamefont {Arya}, \citenamefont {Babbush}, \citenamefont {Bacon}, \citenamefont {Bardin}, \citenamefont {Barends}, \citenamefont {Biswas}, \citenamefont {Boixo}, \citenamefont {Brandao}, \citenamefont {Buell}, \citenamefont {Burkett}, \citenamefont {Chen}, \citenamefont {Chen}, \citenamefont {Chiaro}, \citenamefont {Collins}, \citenamefont {Courtney}, \citenamefont {Dunsworth}, \citenamefont {Farhi}, \citenamefont {Foxen}, \citenamefont {Fowler}, \citenamefont {Gidney}, \citenamefont {Giustina}, \citenamefont {Graff}, \citenamefont {Guerin}, \citenamefont {Habegger}, \citenamefont {Harrigan}, \citenamefont {Hartmann}, \citenamefont {Ho}, \citenamefont {Hoffmann}, \citenamefont {Huang}, \citenamefont {Humble}, \citenamefont {Isakov}, \citenamefont {Jeffrey}, \citenamefont {Jiang}, \citenamefont {Kafri}, \citenamefont {Kechedzhi}, \citenamefont {Kelly}, \citenamefont {Klimov}, \citenamefont {Knysh}, \citenamefont {Korotkov},
  \citenamefont {Kostritsa}, \citenamefont {Landhuis}, \citenamefont {Lindmark}, \citenamefont {Lucero}, \citenamefont {Lyakh}, \citenamefont {Mandrà}, \citenamefont {McClean}, \citenamefont {McEwen}, \citenamefont {Megrant}, \citenamefont {Mi}, \citenamefont {Michielsen}, \citenamefont {Mohseni}, \citenamefont {Mutus}, \citenamefont {Naaman}, \citenamefont {Neeley}, \citenamefont {Neill}, \citenamefont {Niu}, \citenamefont {Ostby}, \citenamefont {Petukhov}, \citenamefont {Platt}, \citenamefont {Quintana}, \citenamefont {Rieffel}, \citenamefont {Roushan}, \citenamefont {Rubin}, \citenamefont {Sank}, \citenamefont {Satzinger}, \citenamefont {Smelyanskiy}, \citenamefont {Sung}, \citenamefont {Trevithick}, \citenamefont {Vainsencher}, \citenamefont {Villalonga}, \citenamefont {White}, \citenamefont {Yao}, \citenamefont {Yeh}, \citenamefont {Zalcman}, \citenamefont {Neven},\ and\ \citenamefont {Martinis}}]{arute2019quantum}%
  \BibitemOpen
  \bibfield  {author} {\bibinfo {author} {\bibfnamefont {F.}~\bibnamefont {Arute}}, \bibinfo {author} {\bibfnamefont {K.}~\bibnamefont {Arya}}, \bibinfo {author} {\bibfnamefont {R.}~\bibnamefont {Babbush}}, \bibinfo {author} {\bibfnamefont {D.}~\bibnamefont {Bacon}}, \bibinfo {author} {\bibfnamefont {J.~C.}\ \bibnamefont {Bardin}}, \bibinfo {author} {\bibfnamefont {R.}~\bibnamefont {Barends}}, \bibinfo {author} {\bibfnamefont {R.}~\bibnamefont {Biswas}}, \bibinfo {author} {\bibfnamefont {S.}~\bibnamefont {Boixo}}, \bibinfo {author} {\bibfnamefont {F.~G. S.~L.}\ \bibnamefont {Brandao}}, \bibinfo {author} {\bibfnamefont {D.~A.}\ \bibnamefont {Buell}}, \bibinfo {author} {\bibfnamefont {B.}~\bibnamefont {Burkett}}, \bibinfo {author} {\bibfnamefont {Y.}~\bibnamefont {Chen}}, \bibinfo {author} {\bibfnamefont {Z.}~\bibnamefont {Chen}}, \bibinfo {author} {\bibfnamefont {B.}~\bibnamefont {Chiaro}}, \bibinfo {author} {\bibfnamefont {R.}~\bibnamefont {Collins}}, \bibinfo {author} {\bibfnamefont {W.}~\bibnamefont
  {Courtney}}, \bibinfo {author} {\bibfnamefont {A.}~\bibnamefont {Dunsworth}}, \bibinfo {author} {\bibfnamefont {E.}~\bibnamefont {Farhi}}, \bibinfo {author} {\bibfnamefont {B.}~\bibnamefont {Foxen}}, \bibinfo {author} {\bibfnamefont {A.}~\bibnamefont {Fowler}}, \bibinfo {author} {\bibfnamefont {C.}~\bibnamefont {Gidney}}, \bibinfo {author} {\bibfnamefont {M.}~\bibnamefont {Giustina}}, \bibinfo {author} {\bibfnamefont {R.}~\bibnamefont {Graff}}, \bibinfo {author} {\bibfnamefont {K.}~\bibnamefont {Guerin}}, \bibinfo {author} {\bibfnamefont {S.}~\bibnamefont {Habegger}}, \bibinfo {author} {\bibfnamefont {M.~P.}\ \bibnamefont {Harrigan}}, \bibinfo {author} {\bibfnamefont {M.~J.}\ \bibnamefont {Hartmann}}, \bibinfo {author} {\bibfnamefont {A.}~\bibnamefont {Ho}}, \bibinfo {author} {\bibfnamefont {M.}~\bibnamefont {Hoffmann}}, \bibinfo {author} {\bibfnamefont {T.}~\bibnamefont {Huang}}, \bibinfo {author} {\bibfnamefont {T.~S.}\ \bibnamefont {Humble}}, \bibinfo {author} {\bibfnamefont {S.~V.}\ \bibnamefont
  {Isakov}}, \bibinfo {author} {\bibfnamefont {E.}~\bibnamefont {Jeffrey}}, \bibinfo {author} {\bibfnamefont {Z.}~\bibnamefont {Jiang}}, \bibinfo {author} {\bibfnamefont {D.}~\bibnamefont {Kafri}}, \bibinfo {author} {\bibfnamefont {K.}~\bibnamefont {Kechedzhi}}, \bibinfo {author} {\bibfnamefont {J.}~\bibnamefont {Kelly}}, \bibinfo {author} {\bibfnamefont {P.~V.}\ \bibnamefont {Klimov}}, \bibinfo {author} {\bibfnamefont {S.}~\bibnamefont {Knysh}}, \bibinfo {author} {\bibfnamefont {A.}~\bibnamefont {Korotkov}}, \bibinfo {author} {\bibfnamefont {F.}~\bibnamefont {Kostritsa}}, \bibinfo {author} {\bibfnamefont {D.}~\bibnamefont {Landhuis}}, \bibinfo {author} {\bibfnamefont {M.}~\bibnamefont {Lindmark}}, \bibinfo {author} {\bibfnamefont {E.}~\bibnamefont {Lucero}}, \bibinfo {author} {\bibfnamefont {D.}~\bibnamefont {Lyakh}}, \bibinfo {author} {\bibfnamefont {S.}~\bibnamefont {Mandrà}}, \bibinfo {author} {\bibfnamefont {J.~R.}\ \bibnamefont {McClean}}, \bibinfo {author} {\bibfnamefont {M.}~\bibnamefont {McEwen}},
  \bibinfo {author} {\bibfnamefont {A.}~\bibnamefont {Megrant}}, \bibinfo {author} {\bibfnamefont {X.}~\bibnamefont {Mi}}, \bibinfo {author} {\bibfnamefont {K.}~\bibnamefont {Michielsen}}, \bibinfo {author} {\bibfnamefont {M.}~\bibnamefont {Mohseni}}, \bibinfo {author} {\bibfnamefont {J.}~\bibnamefont {Mutus}}, \bibinfo {author} {\bibfnamefont {O.}~\bibnamefont {Naaman}}, \bibinfo {author} {\bibfnamefont {M.}~\bibnamefont {Neeley}}, \bibinfo {author} {\bibfnamefont {C.}~\bibnamefont {Neill}}, \bibinfo {author} {\bibfnamefont {M.~Y.}\ \bibnamefont {Niu}}, \bibinfo {author} {\bibfnamefont {E.}~\bibnamefont {Ostby}}, \bibinfo {author} {\bibfnamefont {A.}~\bibnamefont {Petukhov}}, \bibinfo {author} {\bibfnamefont {J.~C.}\ \bibnamefont {Platt}}, \bibinfo {author} {\bibfnamefont {C.}~\bibnamefont {Quintana}}, \bibinfo {author} {\bibfnamefont {E.~G.}\ \bibnamefont {Rieffel}}, \bibinfo {author} {\bibfnamefont {P.}~\bibnamefont {Roushan}}, \bibinfo {author} {\bibfnamefont {N.~C.}\ \bibnamefont {Rubin}}, \bibinfo
  {author} {\bibfnamefont {D.}~\bibnamefont {Sank}}, \bibinfo {author} {\bibfnamefont {K.~J.}\ \bibnamefont {Satzinger}}, \bibinfo {author} {\bibfnamefont {V.}~\bibnamefont {Smelyanskiy}}, \bibinfo {author} {\bibfnamefont {K.~J.}\ \bibnamefont {Sung}}, \bibinfo {author} {\bibfnamefont {M.~D.}\ \bibnamefont {Trevithick}}, \bibinfo {author} {\bibfnamefont {A.}~\bibnamefont {Vainsencher}}, \bibinfo {author} {\bibfnamefont {B.}~\bibnamefont {Villalonga}}, \bibinfo {author} {\bibfnamefont {T.}~\bibnamefont {White}}, \bibinfo {author} {\bibfnamefont {Z.~J.}\ \bibnamefont {Yao}}, \bibinfo {author} {\bibfnamefont {P.}~\bibnamefont {Yeh}}, \bibinfo {author} {\bibfnamefont {A.}~\bibnamefont {Zalcman}}, \bibinfo {author} {\bibfnamefont {H.}~\bibnamefont {Neven}},\ and\ \bibinfo {author} {\bibfnamefont {J.~M.}\ \bibnamefont {Martinis}},\ }\bibfield  {title} {\bibinfo {title} {Quantum supremacy using a programmable superconducting processor},\ }\href {https://www.nature.com/articles/s41586-019-1666-5} {\bibfield
  {journal} {\bibinfo  {journal} {Nature}\ }\textbf {\bibinfo {volume} {574}},\ \bibinfo {pages} {505} (\bibinfo {year} {2019})}\BibitemShut {NoStop}%
\bibitem [{\citenamefont {AI}\ and\ \citenamefont {Collaborators}(2025)}]{google2025}%
  \BibitemOpen
  \bibfield  {author} {\bibinfo {author} {\bibfnamefont {G.~Q.}\ \bibnamefont {AI}}\ and\ \bibinfo {author} {\bibnamefont {Collaborators}},\ }\bibfield  {title} {\bibinfo {title} {Observation of constructive interference at the edge of quantum ergodicity},\ }\href {https://doi.org//10.1038/s41586-025-09526-6} {\bibfield  {journal} {\bibinfo  {journal} {Nature}\ }\textbf {\bibinfo {volume} {646}},\ \bibinfo {pages} {825–830} (\bibinfo {year} {2025})}\BibitemShut {NoStop}%
\bibitem [{\citenamefont {Cerezo}\ \emph {et~al.}(2022)\citenamefont {Cerezo}, \citenamefont {Verdon}, \citenamefont {Huang}, \citenamefont {Cincio},\ and\ \citenamefont {Coles}}]{cerezo_2022_challenges}%
  \BibitemOpen
  \bibfield  {author} {\bibinfo {author} {\bibfnamefont {M.}~\bibnamefont {Cerezo}}, \bibinfo {author} {\bibfnamefont {G.}~\bibnamefont {Verdon}}, \bibinfo {author} {\bibfnamefont {H.-Y.}\ \bibnamefont {Huang}}, \bibinfo {author} {\bibfnamefont {L.}~\bibnamefont {Cincio}},\ and\ \bibinfo {author} {\bibfnamefont {P.~J.}\ \bibnamefont {Coles}},\ }\bibfield  {title} {\bibinfo {title} {Challenges and opportunities in quantum machine learning},\ }\bibfield  {journal} {\bibinfo  {journal} {Nature Computational Science}\ }\href {https://doi.org/10.1038/s43588-022-00311-3} {10.1038/s43588-022-00311-3} (\bibinfo {year} {2022})\BibitemShut {NoStop}%
\bibitem [{\citenamefont {Coles}(2021)}]{coles_2021_seeking}%
  \BibitemOpen
  \bibfield  {author} {\bibinfo {author} {\bibfnamefont {P.~J.}\ \bibnamefont {Coles}},\ }\bibfield  {title} {\bibinfo {title} {Seeking quantum advantage for neural networks},\ }\href {https://doi.org/10.1038/s43588-021-00088-x} {\bibfield  {journal} {\bibinfo  {journal} {Nature Computational Science}\ }\textbf {\bibinfo {volume} {1}},\ \bibinfo {pages} {389} (\bibinfo {year} {2021})}\BibitemShut {NoStop}%
\bibitem [{\citenamefont {Perdomo-Ortiz}\ \emph {et~al.}(2018)\citenamefont {Perdomo-Ortiz}, \citenamefont {Benedetti}, \citenamefont {Realpe-Gómez},\ and\ \citenamefont {Biswas}}]{perdomo_2018}%
  \BibitemOpen
  \bibfield  {author} {\bibinfo {author} {\bibfnamefont {A.}~\bibnamefont {Perdomo-Ortiz}}, \bibinfo {author} {\bibfnamefont {M.}~\bibnamefont {Benedetti}}, \bibinfo {author} {\bibfnamefont {J.}~\bibnamefont {Realpe-Gómez}},\ and\ \bibinfo {author} {\bibfnamefont {R.}~\bibnamefont {Biswas}},\ }\bibfield  {title} {\bibinfo {title} {Opportunities and challenges for quantum-assisted machine learning in near-term quantum computers},\ }\href {https://doi.org/10.1088/2058-9565/aab859} {\bibfield  {journal} {\bibinfo  {journal} {Quantum Science and Technology}\ }\textbf {\bibinfo {volume} {3}},\ \bibinfo {pages} {030502} (\bibinfo {year} {2018})}\BibitemShut {NoStop}%
\bibitem [{\citenamefont {Carlo}\ \emph {et~al.}(2018)\citenamefont {Carlo}, \citenamefont {Mark}, \citenamefont {Davide}, \citenamefont {Massimiliano}, \citenamefont {Andrea}, \citenamefont {Simone},\ and\ \citenamefont {Leonard}}]{ciliberto_2018}%
  \BibitemOpen
  \bibfield  {author} {\bibinfo {author} {\bibfnamefont {C.}~\bibnamefont {Carlo}}, \bibinfo {author} {\bibfnamefont {H.}~\bibnamefont {Mark}}, \bibinfo {author} {\bibfnamefont {I.~A.}\ \bibnamefont {Davide}}, \bibinfo {author} {\bibfnamefont {P.}~\bibnamefont {Massimiliano}}, \bibinfo {author} {\bibfnamefont {R.}~\bibnamefont {Andrea}}, \bibinfo {author} {\bibfnamefont {S.}~\bibnamefont {Simone}},\ and\ \bibinfo {author} {\bibfnamefont {W.}~\bibnamefont {Leonard}},\ }\bibfield  {title} {\bibinfo {title} {Quantum machine learning: a classical perspective},\ }\href {http://doi.org/10.1098/rspa.2017.0551} {\bibfield  {journal} {\bibinfo  {journal} {Proc. R. Soc. A.}\ }\textbf {\bibinfo {volume} {474}} (\bibinfo {year} {2018})}\BibitemShut {NoStop}%
\bibitem [{\citenamefont {Liu}\ \emph {et~al.}(2021)\citenamefont {Liu}, \citenamefont {Arunachalam},\ and\ \citenamefont {Temme}}]{liu_2021}%
  \BibitemOpen
  \bibfield  {author} {\bibinfo {author} {\bibfnamefont {Y.}~\bibnamefont {Liu}}, \bibinfo {author} {\bibfnamefont {S.}~\bibnamefont {Arunachalam}},\ and\ \bibinfo {author} {\bibfnamefont {K.}~\bibnamefont {Temme}},\ }\bibfield  {title} {\bibinfo {title} {A rigorous and robust quantum speed-up in supervised machine learning},\ }\href {https://doi.org/https://doi.org/10.1038/s41567-021-01287-z} {\bibfield  {journal} {\bibinfo  {journal} {Nat. Phys.}\ }\textbf {\bibinfo {volume} {17}},\ \bibinfo {pages} {1013–1017} (\bibinfo {year} {2021})}\BibitemShut {NoStop}%
\bibitem [{\citenamefont {Liao}\ \emph {et~al.}(2021)\citenamefont {Liao}, \citenamefont {Urtasun},\ and\ \citenamefont {Zemel}}]{liao2021}%
  \BibitemOpen
  \bibfield  {author} {\bibinfo {author} {\bibfnamefont {R.}~\bibnamefont {Liao}}, \bibinfo {author} {\bibfnamefont {R.}~\bibnamefont {Urtasun}},\ and\ \bibinfo {author} {\bibfnamefont {R.}~\bibnamefont {Zemel}},\ }\bibfield  {title} {\bibinfo {title} {A {PAC}-{B}ayesian approach to generalization bounds for graph neural networks},\ }in\ \href {https://openreview.net/forum?id=rio73VsWm} {\emph {\bibinfo {booktitle} {International Conference on Learning Representations ({ICLR})}}}\ (\bibinfo {year} {2021})\BibitemShut {NoStop}%
\bibitem [{\citenamefont {Ma}\ \emph {et~al.}(2022)\citenamefont {Ma}, \citenamefont {Rong},\ and\ \citenamefont {Huang}}]{ma2022}%
  \BibitemOpen
  \bibfield  {author} {\bibinfo {author} {\bibfnamefont {H.}~\bibnamefont {Ma}}, \bibinfo {author} {\bibfnamefont {Y.}~\bibnamefont {Rong}},\ and\ \bibinfo {author} {\bibfnamefont {J.}~\bibnamefont {Huang}},\ }\bibfield  {title} {\bibinfo {title} {Graph neural networks: Scalability},\ }in\ \href@noop {} {\emph {\bibinfo {booktitle} {Graph Neural Networks: Foundations, Frontiers, and Applications}}},\ \bibinfo {editor} {edited by\ \bibinfo {editor} {\bibfnamefont {L.}~\bibnamefont {Wu}}, \bibinfo {editor} {\bibfnamefont {P.}~\bibnamefont {Cui}}, \bibinfo {editor} {\bibfnamefont {J.}~\bibnamefont {Pei}},\ and\ \bibinfo {editor} {\bibfnamefont {L.}~\bibnamefont {Zhao}}}\ (\bibinfo  {publisher} {Springer Singapore},\ \bibinfo {address} {Singapore},\ \bibinfo {year} {2022})\ pp.\ \bibinfo {pages} {99--119}\BibitemShut {NoStop}%
\bibitem [{\citenamefont {Kipf}\ and\ \citenamefont {Welling}(2017)}]{welling_GCN_2017}%
  \BibitemOpen
  \bibfield  {author} {\bibinfo {author} {\bibfnamefont {T.~N.}\ \bibnamefont {Kipf}}\ and\ \bibinfo {author} {\bibfnamefont {M.}~\bibnamefont {Welling}},\ }\bibfield  {title} {\bibinfo {title} {Semi-supervised classification with graph convolutional networks},\ }in\ \href {https://openreview.net/forum?id=SJU4ayYgl} {\emph {\bibinfo {booktitle} {International Conference on Learning Representations (ICLR)}}}\ (\bibinfo {year} {2017})\BibitemShut {NoStop}%
\bibitem [{\citenamefont {Hu}\ \emph {et~al.}(2022)\citenamefont {Hu}, \citenamefont {Li}, \citenamefont {Pan}, \citenamefont {Zhou}, \citenamefont {Yang}, \citenamefont {Ding}, \citenamefont {Khan}, \citenamefont {Geng},\ and\ \citenamefont {Jiang}}]{hu_qgnn_2022}%
  \BibitemOpen
  \bibfield  {author} {\bibinfo {author} {\bibfnamefont {Z.}~\bibnamefont {Hu}}, \bibinfo {author} {\bibfnamefont {J.}~\bibnamefont {Li}}, \bibinfo {author} {\bibfnamefont {Z.}~\bibnamefont {Pan}}, \bibinfo {author} {\bibfnamefont {S.}~\bibnamefont {Zhou}}, \bibinfo {author} {\bibfnamefont {L.}~\bibnamefont {Yang}}, \bibinfo {author} {\bibfnamefont {C.}~\bibnamefont {Ding}}, \bibinfo {author} {\bibfnamefont {O.}~\bibnamefont {Khan}}, \bibinfo {author} {\bibfnamefont {T.}~\bibnamefont {Geng}},\ and\ \bibinfo {author} {\bibfnamefont {W.}~\bibnamefont {Jiang}},\ }\bibfield  {title} {\bibinfo {title} {On the design of quantum graph convolutional neural network in the nisq-era and beyond},\ }in\ \href {https://doi.org/10.1109/ICCD56317.2022.00050} {\emph {\bibinfo {booktitle} {2022 IEEE 40th International Conference on Computer Design (ICCD)}}}\ (\bibinfo {year} {2022})\ pp.\ \bibinfo {pages} {290--297}\BibitemShut {NoStop}%
\bibitem [{\citenamefont {Liao}\ \emph {et~al.}(2024)\citenamefont {Liao}, \citenamefont {Zhang},\ and\ \citenamefont {Ferrie}}]{yidong_qgnn_2024}%
  \BibitemOpen
  \bibfield  {author} {\bibinfo {author} {\bibfnamefont {Y.}~\bibnamefont {Liao}}, \bibinfo {author} {\bibfnamefont {X.-M.}\ \bibnamefont {Zhang}},\ and\ \bibinfo {author} {\bibfnamefont {C.}~\bibnamefont {Ferrie}},\ }\bibfield  {title} {\bibinfo {title} {Graph neural networks on quantum computers},\ }\href {https://arxiv.org/abs/2405.17060} {\bibfield  {journal} {\bibinfo  {journal} {arXiv:2405.17060 [quant-ph]}\ } (\bibinfo {year} {2024})}\BibitemShut {NoStop}%
\bibitem [{\citenamefont {Ryu}\ \emph {et~al.}(2023)\citenamefont {Ryu}, \citenamefont {Elala},\ and\ \citenamefont {Rhee}}]{ryu_qgnn_2023}%
  \BibitemOpen
  \bibfield  {author} {\bibinfo {author} {\bibfnamefont {J.-Y.}\ \bibnamefont {Ryu}}, \bibinfo {author} {\bibfnamefont {E.}~\bibnamefont {Elala}},\ and\ \bibinfo {author} {\bibfnamefont {J.-K.~K.}\ \bibnamefont {Rhee}},\ }\bibfield  {title} {\bibinfo {title} {Quantum graph neural network models for materials search},\ }\bibfield  {journal} {\bibinfo  {journal} {Materials}\ }\textbf {\bibinfo {volume} {16}},\ \href {https://doi.org/10.3390/ma16124300} {10.3390/ma16124300} (\bibinfo {year} {2023})\BibitemShut {NoStop}%
\bibitem [{\citenamefont {Zheng}\ \emph {et~al.}(2021)\citenamefont {Zheng}, \citenamefont {Gao},\ and\ \citenamefont {Lü}}]{zheng_qgnn_2021}%
  \BibitemOpen
  \bibfield  {author} {\bibinfo {author} {\bibfnamefont {J.}~\bibnamefont {Zheng}}, \bibinfo {author} {\bibfnamefont {Q.}~\bibnamefont {Gao}},\ and\ \bibinfo {author} {\bibfnamefont {Y.}~\bibnamefont {Lü}},\ }\bibfield  {title} {\bibinfo {title} {Quantum graph convolutional neural networks},\ }in\ \href {https://doi.org/10.23919/CCC52363.2021.9550372} {\emph {\bibinfo {booktitle} {2021 40th Chinese Control Conference (CCC)}}}\ (\bibinfo {organization} {IEEE},\ \bibinfo {year} {2021})\ pp.\ \bibinfo {pages} {6335--6340}\BibitemShut {NoStop}%
\bibitem [{\citenamefont {Zheng}\ \emph {et~al.}(2024)\citenamefont {Zheng}, \citenamefont {Gao}, \citenamefont {Ogorzałek}, \citenamefont {Lü},\ and\ \citenamefont {Deng}}]{zheng_qgnn_2024}%
  \BibitemOpen
  \bibfield  {author} {\bibinfo {author} {\bibfnamefont {J.}~\bibnamefont {Zheng}}, \bibinfo {author} {\bibfnamefont {Q.}~\bibnamefont {Gao}}, \bibinfo {author} {\bibfnamefont {M.}~\bibnamefont {Ogorzałek}}, \bibinfo {author} {\bibfnamefont {J.}~\bibnamefont {Lü}},\ and\ \bibinfo {author} {\bibfnamefont {Y.}~\bibnamefont {Deng}},\ }\bibfield  {title} {\bibinfo {title} {A quantum spatial graph convolutional neural network model on quantum circuits},\ }\href {https://doi.org/10.1109/TNNLS.2024.3382174} {\bibfield  {journal} {\bibinfo  {journal} {IEEE Transactions on Neural Networks and Learning Systems}\ ,\ \bibinfo {pages} {1}} (\bibinfo {year} {2024})}\BibitemShut {NoStop}%
\bibitem [{\citenamefont {Hamilton}\ \emph {et~al.}(2017)\citenamefont {Hamilton}, \citenamefont {Ying},\ and\ \citenamefont {Leskovec}}]{leskovec_inductive_representation_2018}%
  \BibitemOpen
  \bibfield  {author} {\bibinfo {author} {\bibfnamefont {W.~L.}\ \bibnamefont {Hamilton}}, \bibinfo {author} {\bibfnamefont {R.}~\bibnamefont {Ying}},\ and\ \bibinfo {author} {\bibfnamefont {J.}~\bibnamefont {Leskovec}},\ }\bibfield  {title} {\bibinfo {title} {Inductive representation learning on large graphs},\ }in\ \href@noop {} {\emph {\bibinfo {booktitle} {Advances in Neural Information Processing Systems (NeurIPS)}}},\ Vol.~\bibinfo {volume} {30}\ (\bibinfo {year} {2017})\BibitemShut {NoStop}%
\bibitem [{\citenamefont {Cong}\ \emph {et~al.}(2019)\citenamefont {Cong}, \citenamefont {Choi},\ and\ \citenamefont {Lukin}}]{lukin_qcnn_2019}%
  \BibitemOpen
  \bibfield  {author} {\bibinfo {author} {\bibfnamefont {I.}~\bibnamefont {Cong}}, \bibinfo {author} {\bibfnamefont {S.}~\bibnamefont {Choi}},\ and\ \bibinfo {author} {\bibfnamefont {M.~D.}\ \bibnamefont {Lukin}},\ }\bibfield  {title} {\bibinfo {title} {Quantum convolutional neural networks},\ }\href {https://doi.org/doi.org/10.1038/s41567-019-0648-8} {\bibfield  {journal} {\bibinfo  {journal} {Nat. Phys.}\ }\textbf {\bibinfo {volume} {15}},\ \bibinfo {pages} {1273–1278} (\bibinfo {year} {2019})}\BibitemShut {NoStop}%
\bibitem [{\citenamefont {Fey}\ and\ \citenamefont {Lenssen}(2019)}]{pyg_code}%
  \BibitemOpen
  \bibfield  {author} {\bibinfo {author} {\bibfnamefont {M.}~\bibnamefont {Fey}}\ and\ \bibinfo {author} {\bibfnamefont {J.~E.}\ \bibnamefont {Lenssen}},\ }\href {https://github.com/pyg-team/pytorch_geometric} {\bibinfo {title} {{PyTorch Geometric: A Library for Graph Neural Networks}}} (\bibinfo {year} {2019})\BibitemShut {NoStop}%
\bibitem [{\citenamefont {Wu}\ \emph {et~al.}(2018)\citenamefont {Wu}, \citenamefont {Ramsundar}, \citenamefont {Feinberg}, \citenamefont {Gomes}, \citenamefont {Geniesse}, \citenamefont {Pappu}, \citenamefont {Leswing},\ and\ \citenamefont {Pande}}]{moleculeNet_2017}%
  \BibitemOpen
  \bibfield  {author} {\bibinfo {author} {\bibfnamefont {Z.}~\bibnamefont {Wu}}, \bibinfo {author} {\bibfnamefont {B.}~\bibnamefont {Ramsundar}}, \bibinfo {author} {\bibfnamefont {E.~N.}\ \bibnamefont {Feinberg}}, \bibinfo {author} {\bibfnamefont {J.}~\bibnamefont {Gomes}}, \bibinfo {author} {\bibfnamefont {C.}~\bibnamefont {Geniesse}}, \bibinfo {author} {\bibfnamefont {A.~S.}\ \bibnamefont {Pappu}}, \bibinfo {author} {\bibfnamefont {K.}~\bibnamefont {Leswing}},\ and\ \bibinfo {author} {\bibfnamefont {V.}~\bibnamefont {Pande}},\ }\bibfield  {title} {\bibinfo {title} {Moleculenet: A benchmark for molecular machine learning},\ }\href {https://doi.org/10.1039/C7SC02664A} {\bibfield  {journal} {\bibinfo  {journal} {Chem. Sci.}\ }\textbf {\bibinfo {volume} {9}},\ \bibinfo {pages} {513} (\bibinfo {year} {2018})}\BibitemShut {NoStop}%
\bibitem [{\citenamefont {Anschuetz}\ and\ \citenamefont {Kiani}(2022)}]{anschuetz_2022_beyond}%
  \BibitemOpen
  \bibfield  {author} {\bibinfo {author} {\bibfnamefont {E.~R.}\ \bibnamefont {Anschuetz}}\ and\ \bibinfo {author} {\bibfnamefont {B.~T.}\ \bibnamefont {Kiani}},\ }\bibfield  {title} {\bibinfo {title} {Quantum variational algorithms are swamped with traps},\ }\href {https://doi.org/doi.org/10.1038/s41467-022-35364-5} {\bibfield  {journal} {\bibinfo  {journal} {Nat Commun}\ }\textbf {\bibinfo {volume} {13}},\ \bibinfo {pages} {7760} (\bibinfo {year} {2022})}\BibitemShut {NoStop}%
\bibitem [{\citenamefont {Ragone}\ \emph {et~al.}(2023)\citenamefont {Ragone}, \citenamefont {Bakalov}, \citenamefont {Sauvage}, \citenamefont {Kemper}, \citenamefont {Marrero}, \citenamefont {Larocca},\ and\ \citenamefont {Cerezo}}]{ragone_2023_lie}%
  \BibitemOpen
  \bibfield  {author} {\bibinfo {author} {\bibfnamefont {M.}~\bibnamefont {Ragone}}, \bibinfo {author} {\bibfnamefont {B.~N.}\ \bibnamefont {Bakalov}}, \bibinfo {author} {\bibfnamefont {F.}~\bibnamefont {Sauvage}}, \bibinfo {author} {\bibfnamefont {A.~F.}\ \bibnamefont {Kemper}}, \bibinfo {author} {\bibfnamefont {C.~O.}\ \bibnamefont {Marrero}}, \bibinfo {author} {\bibfnamefont {M.}~\bibnamefont {Larocca}},\ and\ \bibinfo {author} {\bibfnamefont {M.}~\bibnamefont {Cerezo}},\ }\bibfield  {title} {\bibinfo {title} {A lie algebraic theory of barren plateaus for deep parameterized quantum circuits},\ }\bibfield  {journal} {\bibinfo  {journal} {Nat Commun}\ }\textbf {\bibinfo {volume} {15}},\ \href {https://doi.org/10.1038/s41467-024-49909-3} {10.1038/s41467-024-49909-3} (\bibinfo {year} {2023})\BibitemShut {NoStop}%
\bibitem [{\citenamefont {Larocca}\ \emph {et~al.}(2025)\citenamefont {Larocca}, \citenamefont {Thanasilp}, \citenamefont {Wang}, \citenamefont {Sharma}, \citenamefont {Biamonte}, \citenamefont {Coles}, \citenamefont {Cincio}, \citenamefont {McClean}, \citenamefont {Holmes},\ and\ \citenamefont {Cerezo}}]{larocca_2024_review}%
  \BibitemOpen
  \bibfield  {author} {\bibinfo {author} {\bibfnamefont {M.}~\bibnamefont {Larocca}}, \bibinfo {author} {\bibfnamefont {S.}~\bibnamefont {Thanasilp}}, \bibinfo {author} {\bibfnamefont {S.}~\bibnamefont {Wang}}, \bibinfo {author} {\bibfnamefont {K.}~\bibnamefont {Sharma}}, \bibinfo {author} {\bibfnamefont {J.}~\bibnamefont {Biamonte}}, \bibinfo {author} {\bibfnamefont {P.~J.}\ \bibnamefont {Coles}}, \bibinfo {author} {\bibfnamefont {L.}~\bibnamefont {Cincio}}, \bibinfo {author} {\bibfnamefont {J.~R.}\ \bibnamefont {McClean}}, \bibinfo {author} {\bibfnamefont {Z.}~\bibnamefont {Holmes}},\ and\ \bibinfo {author} {\bibfnamefont {M.}~\bibnamefont {Cerezo}},\ }\bibfield  {title} {\bibinfo {title} {A review of barren plateaus in variational quantum computing},\ }\href {https://doi.org/10.1038/s42254-025-00813-9} {\bibfield  {journal} {\bibinfo  {journal} {Nat Rev Phys}\ }\textbf {\bibinfo {volume} {7}},\ \bibinfo {pages} {174–189} (\bibinfo {year} {2025})}\BibitemShut {NoStop}%
\bibitem [{\citenamefont {Arrasmith}\ \emph {et~al.}(2022)\citenamefont {Arrasmith}, \citenamefont {Holmes}, \citenamefont {Cerezo},\ and\ \citenamefont {Coles}}]{arrasmith_2021_equivalence}%
  \BibitemOpen
  \bibfield  {author} {\bibinfo {author} {\bibfnamefont {A.}~\bibnamefont {Arrasmith}}, \bibinfo {author} {\bibfnamefont {Z.}~\bibnamefont {Holmes}}, \bibinfo {author} {\bibfnamefont {M.}~\bibnamefont {Cerezo}},\ and\ \bibinfo {author} {\bibfnamefont {P.~J.}\ \bibnamefont {Coles}},\ }\bibfield  {title} {\bibinfo {title} {Equivalence of quantum barren plateaus to cost concentration and narrow gorges},\ }\href {https://doi.org/10.1088/2058-9565/ac7d06} {\bibfield  {journal} {\bibinfo  {journal} {Quantum Science and Technology}\ }\textbf {\bibinfo {volume} {7}},\ \bibinfo {pages} {045015} (\bibinfo {year} {2022})}\BibitemShut {NoStop}%
\bibitem [{\citenamefont {Verdon}\ \emph {et~al.}(2019)\citenamefont {Verdon}, \citenamefont {McCourt}, \citenamefont {Luzhnica}, \citenamefont {Singh}, \citenamefont {Leichenauer},\ and\ \citenamefont {Hidary}}]{verdon_2019_quantumgraph}%
  \BibitemOpen
  \bibfield  {author} {\bibinfo {author} {\bibfnamefont {G.}~\bibnamefont {Verdon}}, \bibinfo {author} {\bibfnamefont {T.}~\bibnamefont {McCourt}}, \bibinfo {author} {\bibfnamefont {E.}~\bibnamefont {Luzhnica}}, \bibinfo {author} {\bibfnamefont {V.}~\bibnamefont {Singh}}, \bibinfo {author} {\bibfnamefont {S.}~\bibnamefont {Leichenauer}},\ and\ \bibinfo {author} {\bibfnamefont {J.}~\bibnamefont {Hidary}},\ }\bibfield  {title} {\bibinfo {title} {Quantum graph neural networks},\ }\href {https://arxiv.org/abs/1909.12264} {\bibfield  {journal} {\bibinfo  {journal} {arXiv:1909.12264 [quant-ph]}\ } (\bibinfo {year} {2019})}\BibitemShut {NoStop}%
\bibitem [{\citenamefont {Thabet}\ \emph {et~al.}(2024)\citenamefont {Thabet}, \citenamefont {Djellabi}, \citenamefont {Sokolov}, \citenamefont {Kasture}, \citenamefont {Henry},\ and\ \citenamefont {Henriet}}]{thabet_2024}%
  \BibitemOpen
  \bibfield  {author} {\bibinfo {author} {\bibfnamefont {S.}~\bibnamefont {Thabet}}, \bibinfo {author} {\bibfnamefont {M.}~\bibnamefont {Djellabi}}, \bibinfo {author} {\bibfnamefont {I.~O.}\ \bibnamefont {Sokolov}}, \bibinfo {author} {\bibfnamefont {S.}~\bibnamefont {Kasture}}, \bibinfo {author} {\bibfnamefont {L.-P.}\ \bibnamefont {Henry}},\ and\ \bibinfo {author} {\bibfnamefont {L.}~\bibnamefont {Henriet}},\ }\bibfield  {title} {\bibinfo {title} {Quantum positional encodings for graph neural networks},\ }in\ \href {https://proceedings.mlr.press/v235/thabet24a.html} {\emph {\bibinfo {booktitle} {Proceedings of the 41st International Conference on Machine Learning (ICML)}}},\ Vol.\ \bibinfo {volume} {235}\ (\bibinfo  {publisher} {PMLR},\ \bibinfo {year} {2024})\ pp.\ \bibinfo {pages} {47965--47996}\BibitemShut {NoStop}%
\bibitem [{\citenamefont {Beer}\ \emph {et~al.}(2023)\citenamefont {Beer}, \citenamefont {Khosla}, \citenamefont {K\"ohler}, \citenamefont {Osborne},\ and\ \citenamefont {Zhao}}]{beer_2021_quantum}%
  \BibitemOpen
  \bibfield  {author} {\bibinfo {author} {\bibfnamefont {K.}~\bibnamefont {Beer}}, \bibinfo {author} {\bibfnamefont {M.}~\bibnamefont {Khosla}}, \bibinfo {author} {\bibfnamefont {J.}~\bibnamefont {K\"ohler}}, \bibinfo {author} {\bibfnamefont {T.~J.}\ \bibnamefont {Osborne}},\ and\ \bibinfo {author} {\bibfnamefont {T.}~\bibnamefont {Zhao}},\ }\bibfield  {title} {\bibinfo {title} {Quantum machine learning of graph-structured data},\ }\href {https://doi.org/10.1103/PhysRevA.108.012410} {\bibfield  {journal} {\bibinfo  {journal} {Phys. Rev. A}\ }\textbf {\bibinfo {volume} {108}},\ \bibinfo {pages} {012410} (\bibinfo {year} {2023})}\BibitemShut {NoStop}%
\bibitem [{\citenamefont {Skolik}\ \emph {et~al.}(2023)\citenamefont {Skolik}, \citenamefont {Cattelan}, \citenamefont {Yarkoni}, \citenamefont {B{\"a}ck},\ and\ \citenamefont {Dunjko}}]{skolik_2023_equivariant}%
  \BibitemOpen
  \bibfield  {author} {\bibinfo {author} {\bibfnamefont {A.}~\bibnamefont {Skolik}}, \bibinfo {author} {\bibfnamefont {M.}~\bibnamefont {Cattelan}}, \bibinfo {author} {\bibfnamefont {S.}~\bibnamefont {Yarkoni}}, \bibinfo {author} {\bibfnamefont {T.}~\bibnamefont {B{\"a}ck}},\ and\ \bibinfo {author} {\bibfnamefont {V.}~\bibnamefont {Dunjko}},\ }\bibfield  {title} {\bibinfo {title} {Equivariant quantum circuits for learning on weighted graphs},\ }\href@noop {} {\bibfield  {journal} {\bibinfo  {journal} {npj Quantum Information}\ }\textbf {\bibinfo {volume} {9}},\ \bibinfo {pages} {47} (\bibinfo {year} {2023})}\BibitemShut {NoStop}%
\bibitem [{\citenamefont {Ai}\ \emph {et~al.}(2024)\citenamefont {Ai}, \citenamefont {Zhang}, \citenamefont {Sun}, \citenamefont {Yan},\ and\ \citenamefont {Hancock}}]{xing_qgnn_2024}%
  \BibitemOpen
  \bibfield  {author} {\bibinfo {author} {\bibfnamefont {X.}~\bibnamefont {Ai}}, \bibinfo {author} {\bibfnamefont {Z.}~\bibnamefont {Zhang}}, \bibinfo {author} {\bibfnamefont {L.}~\bibnamefont {Sun}}, \bibinfo {author} {\bibfnamefont {J.}~\bibnamefont {Yan}},\ and\ \bibinfo {author} {\bibfnamefont {E.}~\bibnamefont {Hancock}},\ }\bibfield  {title} {\bibinfo {title} {Towards quantum graph neural networks: An ego-graph learning approach},\ }\href@noop {} {\bibfield  {journal} {\bibinfo  {journal} {arXiv:2201.05158 [quant-ph]}\ } (\bibinfo {year} {2024})}\BibitemShut {NoStop}%
\bibitem [{\citenamefont {Zhou}\ \emph {et~al.}(2022)\citenamefont {Zhou}, \citenamefont {Zhang}, \citenamefont {Chen}, \citenamefont {Chen}, \citenamefont {Huang}, \citenamefont {Wang},\ and\ \citenamefont {Hu}}]{kaixiong_qgnn_2022}%
  \BibitemOpen
  \bibfield  {author} {\bibinfo {author} {\bibfnamefont {K.}~\bibnamefont {Zhou}}, \bibinfo {author} {\bibfnamefont {Z.}~\bibnamefont {Zhang}}, \bibinfo {author} {\bibfnamefont {S.}~\bibnamefont {Chen}}, \bibinfo {author} {\bibfnamefont {T.}~\bibnamefont {Chen}}, \bibinfo {author} {\bibfnamefont {X.}~\bibnamefont {Huang}}, \bibinfo {author} {\bibfnamefont {Z.}~\bibnamefont {Wang}},\ and\ \bibinfo {author} {\bibfnamefont {X.}~\bibnamefont {Hu}},\ }\bibfield  {title} {\bibinfo {title} {Quangcn: Noise-adaptive training for robust quantum graph convolutional networks},\ }\href@noop {} {\bibfield  {journal} {\bibinfo  {journal} {arXiv:2211.07379 [quant-ph]}\ } (\bibinfo {year} {2022})}\BibitemShut {NoStop}%
\bibitem [{\citenamefont {T{\"u}ys{\"u}z}\ \emph {et~al.}(2021)\citenamefont {T{\"u}ys{\"u}z}, \citenamefont {Rieger}, \citenamefont {Novotny}, \citenamefont {Demirk{\"o}z}, \citenamefont {Dobos}, \citenamefont {Potamianos}, \citenamefont {Vallecorsa}, \citenamefont {Vlimant},\ and\ \citenamefont {Forster}}]{tuysuz_2021_hybrid}%
  \BibitemOpen
  \bibfield  {author} {\bibinfo {author} {\bibfnamefont {C.}~\bibnamefont {T{\"u}ys{\"u}z}}, \bibinfo {author} {\bibfnamefont {C.}~\bibnamefont {Rieger}}, \bibinfo {author} {\bibfnamefont {K.}~\bibnamefont {Novotny}}, \bibinfo {author} {\bibfnamefont {B.}~\bibnamefont {Demirk{\"o}z}}, \bibinfo {author} {\bibfnamefont {D.}~\bibnamefont {Dobos}}, \bibinfo {author} {\bibfnamefont {K.}~\bibnamefont {Potamianos}}, \bibinfo {author} {\bibfnamefont {S.}~\bibnamefont {Vallecorsa}}, \bibinfo {author} {\bibfnamefont {J.-R.}\ \bibnamefont {Vlimant}},\ and\ \bibinfo {author} {\bibfnamefont {R.}~\bibnamefont {Forster}},\ }\bibfield  {title} {\bibinfo {title} {Hybrid quantum classical graph neural networks for particle track reconstruction},\ }\href@noop {} {\bibfield  {journal} {\bibinfo  {journal} {Quantum Machine Intelligence}\ }\textbf {\bibinfo {volume} {3}},\ \bibinfo {pages} {1} (\bibinfo {year} {2021})}\BibitemShut {NoStop}%
\bibitem [{\citenamefont {Mernyei}\ \emph {et~al.}(2022)\citenamefont {Mernyei}, \citenamefont {Meichanetzidis},\ and\ \citenamefont {Ceylan}}]{mernyei_2022_equivariant}%
  \BibitemOpen
  \bibfield  {author} {\bibinfo {author} {\bibfnamefont {P.}~\bibnamefont {Mernyei}}, \bibinfo {author} {\bibfnamefont {K.}~\bibnamefont {Meichanetzidis}},\ and\ \bibinfo {author} {\bibfnamefont {I.~I.}\ \bibnamefont {Ceylan}},\ }\bibfield  {title} {\bibinfo {title} {Equivariant quantum graph circuits},\ }in\ \href {https://proceedings.mlr.press/v162/mernyei22a.html} {\emph {\bibinfo {booktitle} {International Conference on Machine Learning}}}\ (\bibinfo {organization} {PMLR},\ \bibinfo {year} {2022})\ pp.\ \bibinfo {pages} {15401--15420}\BibitemShut {NoStop}%
\bibitem [{\citenamefont {Kyriienko}\ \emph {et~al.}(2021)\citenamefont {Kyriienko}, \citenamefont {Paine},\ and\ \citenamefont {Elfving}}]{kyriienko_2021_solving}%
  \BibitemOpen
  \bibfield  {author} {\bibinfo {author} {\bibfnamefont {O.}~\bibnamefont {Kyriienko}}, \bibinfo {author} {\bibfnamefont {A.~E.}\ \bibnamefont {Paine}},\ and\ \bibinfo {author} {\bibfnamefont {V.~E.}\ \bibnamefont {Elfving}},\ }\bibfield  {title} {\bibinfo {title} {Solving nonlinear differential equations with differentiable quantum circuits},\ }\href {https://doi.org/10.1103/PhysRevA.103.052416} {\bibfield  {journal} {\bibinfo  {journal} {Phys. Rev. A}\ }\textbf {\bibinfo {volume} {103}},\ \bibinfo {pages} {052416} (\bibinfo {year} {2021})}\BibitemShut {NoStop}%
\bibitem [{\citenamefont {Hamilton}\ \emph {et~al.}(2018)\citenamefont {Hamilton}, \citenamefont {Bajaj}, \citenamefont {Zitnik}, \citenamefont {Jurafsky},\ and\ \citenamefont {Leskovec}}]{leskovec_embedding_2019}%
  \BibitemOpen
  \bibfield  {author} {\bibinfo {author} {\bibfnamefont {W.~L.}\ \bibnamefont {Hamilton}}, \bibinfo {author} {\bibfnamefont {P.}~\bibnamefont {Bajaj}}, \bibinfo {author} {\bibfnamefont {M.}~\bibnamefont {Zitnik}}, \bibinfo {author} {\bibfnamefont {D.}~\bibnamefont {Jurafsky}},\ and\ \bibinfo {author} {\bibfnamefont {J.}~\bibnamefont {Leskovec}},\ }\bibfield  {title} {\bibinfo {title} {Embedding logical queries on knowledge graphs},\ }in\ \href {https://proceedings.neurips.cc/paper/2018/hash/ef50c335cca9f340bde656363ebd02fd-Abstract.html} {\emph {\bibinfo {booktitle} {Advances in Neural Information Processing Systems (NeurIPS)}}},\ Vol.~\bibinfo {volume} {31}\ (\bibinfo {year} {2018})\ pp.\ \bibinfo {pages} {2030--2041}\BibitemShut {NoStop}%
\bibitem [{\citenamefont {Hochreiter}\ and\ \citenamefont {Schmidhuber}(1997)}]{hochreiter_1997_long}%
  \BibitemOpen
  \bibfield  {author} {\bibinfo {author} {\bibfnamefont {S.}~\bibnamefont {Hochreiter}}\ and\ \bibinfo {author} {\bibfnamefont {J.}~\bibnamefont {Schmidhuber}},\ }\bibfield  {title} {\bibinfo {title} {Long short-term memory},\ }\href@noop {} {\bibfield  {journal} {\bibinfo  {journal} {Neural computation}\ }\textbf {\bibinfo {volume} {9}},\ \bibinfo {pages} {1735} (\bibinfo {year} {1997})}\BibitemShut {NoStop}%
\bibitem [{\citenamefont {Qi}\ \emph {et~al.}(2017)\citenamefont {Qi}, \citenamefont {Su}, \citenamefont {Mo},\ and\ \citenamefont {Guibas}}]{qi_2016_pointnet}%
  \BibitemOpen
  \bibfield  {author} {\bibinfo {author} {\bibfnamefont {C.~R.}\ \bibnamefont {Qi}}, \bibinfo {author} {\bibfnamefont {H.}~\bibnamefont {Su}}, \bibinfo {author} {\bibfnamefont {K.}~\bibnamefont {Mo}},\ and\ \bibinfo {author} {\bibfnamefont {L.~J.}\ \bibnamefont {Guibas}},\ }\bibfield  {title} {\bibinfo {title} {Pointnet: Deep learning on point sets for 3d classification and segmentation},\ }\href {https://arxiv.org/abs/1612.00593} {\bibfield  {journal} {\bibinfo  {journal} {arXiv:1612.00593 [cs.CV]}\ } (\bibinfo {year} {2017})}\BibitemShut {NoStop}%
\bibitem [{\citenamefont {Veli{\v{c}}kovi{\'c}}\ \emph {et~al.}(2018)\citenamefont {Veli{\v{c}}kovi{\'c}}, \citenamefont {Cucurull}, \citenamefont {Casanova}, \citenamefont {Romero}, \citenamefont {Li{\`o}},\ and\ \citenamefont {Bengio}}]{petar_gats_2018}%
  \BibitemOpen
  \bibfield  {author} {\bibinfo {author} {\bibfnamefont {P.}~\bibnamefont {Veli{\v{c}}kovi{\'c}}}, \bibinfo {author} {\bibfnamefont {G.}~\bibnamefont {Cucurull}}, \bibinfo {author} {\bibfnamefont {A.}~\bibnamefont {Casanova}}, \bibinfo {author} {\bibfnamefont {A.}~\bibnamefont {Romero}}, \bibinfo {author} {\bibfnamefont {P.}~\bibnamefont {Li{\`o}}},\ and\ \bibinfo {author} {\bibfnamefont {Y.}~\bibnamefont {Bengio}},\ }\bibfield  {title} {\bibinfo {title} {Graph attention networks},\ }in\ \href {https://openreview.net/forum?id=rJXMpikCZ} {\emph {\bibinfo {booktitle} {International Conference on Learning Representations (ICLR)}}}\ (\bibinfo {year} {2018})\BibitemShut {NoStop}%
\bibitem [{\citenamefont {Schuld}\ \emph {et~al.}(2021)\citenamefont {Schuld}, \citenamefont {Sweke},\ and\ \citenamefont {Meyer}}]{schuld_2021}%
  \BibitemOpen
  \bibfield  {author} {\bibinfo {author} {\bibfnamefont {M.}~\bibnamefont {Schuld}}, \bibinfo {author} {\bibfnamefont {R.}~\bibnamefont {Sweke}},\ and\ \bibinfo {author} {\bibfnamefont {J.~J.}\ \bibnamefont {Meyer}},\ }\bibfield  {title} {\bibinfo {title} {Effect of data encoding on the expressive power of variational quantum-machine-learning models},\ }\href {https://doi.org/10.1103/PhysRevA.103.032430} {\bibfield  {journal} {\bibinfo  {journal} {Phys. Rev. A}\ }\textbf {\bibinfo {volume} {103}},\ \bibinfo {pages} {032430} (\bibinfo {year} {2021})}\BibitemShut {NoStop}%
\bibitem [{\citenamefont {Vatan}\ and\ \citenamefont {Williams}(2004)}]{vatan_2004_optimal}%
  \BibitemOpen
  \bibfield  {author} {\bibinfo {author} {\bibfnamefont {F.}~\bibnamefont {Vatan}}\ and\ \bibinfo {author} {\bibfnamefont {C.}~\bibnamefont {Williams}},\ }\bibfield  {title} {\bibinfo {title} {Optimal quantum circuits for general two-qubit gates},\ }\href {https://doi.org/10.1103/PhysRevA.69.032315} {\bibfield  {journal} {\bibinfo  {journal} {Phys. Rev. A}\ }\textbf {\bibinfo {volume} {69}},\ \bibinfo {pages} {032315} (\bibinfo {year} {2004})}\BibitemShut {NoStop}%
\bibitem [{\citenamefont {Larocca}\ \emph {et~al.}(2022)\citenamefont {Larocca}, \citenamefont {Sauvage}, \citenamefont {Sbahi}, \citenamefont {Verdon}, \citenamefont {Coles},\ and\ \citenamefont {Cerezo}}]{larocca_2022_group}%
  \BibitemOpen
  \bibfield  {author} {\bibinfo {author} {\bibfnamefont {M.}~\bibnamefont {Larocca}}, \bibinfo {author} {\bibfnamefont {F.}~\bibnamefont {Sauvage}}, \bibinfo {author} {\bibfnamefont {F.~M.}\ \bibnamefont {Sbahi}}, \bibinfo {author} {\bibfnamefont {G.}~\bibnamefont {Verdon}}, \bibinfo {author} {\bibfnamefont {P.~J.}\ \bibnamefont {Coles}},\ and\ \bibinfo {author} {\bibfnamefont {M.}~\bibnamefont {Cerezo}},\ }\bibfield  {title} {\bibinfo {title} {Group-invariant quantum machine learning},\ }\href {https://doi.org/10.1103/PRXQuantum.3.030341} {\bibfield  {journal} {\bibinfo  {journal} {PRX Quantum}\ }\textbf {\bibinfo {volume} {3}},\ \bibinfo {pages} {030341} (\bibinfo {year} {2022})}\BibitemShut {NoStop}%
\bibitem [{\citenamefont {Chinzei}\ \emph {et~al.}(2024)\citenamefont {Chinzei}, \citenamefont {Tran}, \citenamefont {Maruyama}, \citenamefont {Oshima},\ and\ \citenamefont {Sato}}]{chinzei_2023}%
  \BibitemOpen
  \bibfield  {author} {\bibinfo {author} {\bibfnamefont {K.}~\bibnamefont {Chinzei}}, \bibinfo {author} {\bibfnamefont {Q.~H.}\ \bibnamefont {Tran}}, \bibinfo {author} {\bibfnamefont {K.}~\bibnamefont {Maruyama}}, \bibinfo {author} {\bibfnamefont {H.}~\bibnamefont {Oshima}},\ and\ \bibinfo {author} {\bibfnamefont {S.}~\bibnamefont {Sato}},\ }\bibfield  {title} {\bibinfo {title} {Splitting and parallelizing of quantum convolutional neural networks for learning translationally symmetric data},\ }\href {https://doi.org/10.1103/PhysRevResearch.6.023042} {\bibfield  {journal} {\bibinfo  {journal} {Phys. Rev. Res.}\ }\textbf {\bibinfo {volume} {6}},\ \bibinfo {pages} {023042} (\bibinfo {year} {2024})}\BibitemShut {NoStop}%
\bibitem [{\citenamefont {Nguyen}\ \emph {et~al.}(2024)\citenamefont {Nguyen}, \citenamefont {Schatzki}, \citenamefont {Braccia}, \citenamefont {Ragone}, \citenamefont {Coles}, \citenamefont {Sauvage}, \citenamefont {Larocca},\ and\ \citenamefont {Cerezo}}]{cerezo_2022_theory_equivariant}%
  \BibitemOpen
  \bibfield  {author} {\bibinfo {author} {\bibfnamefont {Q.~T.}\ \bibnamefont {Nguyen}}, \bibinfo {author} {\bibfnamefont {L.}~\bibnamefont {Schatzki}}, \bibinfo {author} {\bibfnamefont {P.}~\bibnamefont {Braccia}}, \bibinfo {author} {\bibfnamefont {M.}~\bibnamefont {Ragone}}, \bibinfo {author} {\bibfnamefont {P.~J.}\ \bibnamefont {Coles}}, \bibinfo {author} {\bibfnamefont {F.}~\bibnamefont {Sauvage}}, \bibinfo {author} {\bibfnamefont {M.}~\bibnamefont {Larocca}},\ and\ \bibinfo {author} {\bibfnamefont {M.}~\bibnamefont {Cerezo}},\ }\bibfield  {title} {\bibinfo {title} {Theory for equivariant quantum neural networks},\ }\href {https://doi.org/10.1103/PRXQuantum.5.020328} {\bibfield  {journal} {\bibinfo  {journal} {PRX Quantum}\ }\textbf {\bibinfo {volume} {5}},\ \bibinfo {pages} {020328} (\bibinfo {year} {2024})}\BibitemShut {NoStop}%
\bibitem [{\citenamefont {Zeng}\ and\ \citenamefont {Zhou}(2016)}]{zeng_2016}%
  \BibitemOpen
  \bibfield  {author} {\bibinfo {author} {\bibfnamefont {B.}~\bibnamefont {Zeng}}\ and\ \bibinfo {author} {\bibfnamefont {D.~L.}\ \bibnamefont {Zhou}},\ }\bibfield  {title} {\bibinfo {title} {Topological and error-correcting properties for symmetry-protected topological order},\ }\href {https://doi.org/10.1209/0295-5075/113/56001} {\bibfield  {journal} {\bibinfo  {journal} {Europhysics Letters}\ }\textbf {\bibinfo {volume} {113}},\ \bibinfo {pages} {56001} (\bibinfo {year} {2016})}\BibitemShut {NoStop}%
\bibitem [{\citenamefont {Nielsen}\ and\ \citenamefont {Chuang}(2010)}]{nielsen}%
  \BibitemOpen
  \bibfield  {author} {\bibinfo {author} {\bibfnamefont {M.~A.}\ \bibnamefont {Nielsen}}\ and\ \bibinfo {author} {\bibfnamefont {I.~L.}\ \bibnamefont {Chuang}},\ }\href@noop {} {\emph {\bibinfo {title} {Quantum Computation and Quantum Information}}}\ (\bibinfo  {publisher} {Cambridge University Press},\ \bibinfo {year} {2010})\BibitemShut {NoStop}%
\bibitem [{\citenamefont {Kingma}\ and\ \citenamefont {Ba}(2014)}]{adam_2014}%
  \BibitemOpen
  \bibfield  {author} {\bibinfo {author} {\bibfnamefont {D.~P.}\ \bibnamefont {Kingma}}\ and\ \bibinfo {author} {\bibfnamefont {J.}~\bibnamefont {Ba}},\ }\bibfield  {title} {\bibinfo {title} {Adam: A method for stochastic optimization},\ }\href {https://arxiv.org/abs/1412.6980} {\bibfield  {journal} {\bibinfo  {journal} {arXiv:1412.6980 [cs.LG]}\ } (\bibinfo {year} {2014})}\BibitemShut {NoStop}%
\bibitem [{\citenamefont {Ansel}\ \emph {et~al.}(2024)\citenamefont {Ansel}, \citenamefont {Yang}, \citenamefont {He}, \citenamefont {Gimelshein}, \citenamefont {Jain}, \citenamefont {Voznesensky}, \citenamefont {Bao}, \citenamefont {Bell}, \citenamefont {Berard}, \citenamefont {Burovski}, \citenamefont {Chauhan}, \citenamefont {Chourdia}, \citenamefont {Constable}, \citenamefont {Desmaison}, \citenamefont {DeVito}, \citenamefont {Ellison}, \citenamefont {Feng}, \citenamefont {Gong}, \citenamefont {Gschwind}, \citenamefont {Hirsh}, \citenamefont {Huang}, \citenamefont {Kalambarkar}, \citenamefont {Kirsch}, \citenamefont {Lazos}, \citenamefont {Lezcano}, \citenamefont {Liang}, \citenamefont {Liang}, \citenamefont {Lu}, \citenamefont {Luk}, \citenamefont {Maher}, \citenamefont {Pan}, \citenamefont {Puhrsch}, \citenamefont {Reso}, \citenamefont {Saroufim}, \citenamefont {Siraichi}, \citenamefont {Suk}, \citenamefont {Suo}, \citenamefont {Tillet}, \citenamefont {Wang}, \citenamefont {Wang}, \citenamefont {Wen},
  \citenamefont {Zhang}, \citenamefont {Zhao}, \citenamefont {Zhou}, \citenamefont {Zou}, \citenamefont {Mathews}, \citenamefont {Chanan}, \citenamefont {Wu},\ and\ \citenamefont {Chintala}}]{pytorch}%
  \BibitemOpen
  \bibfield  {author} {\bibinfo {author} {\bibfnamefont {J.}~\bibnamefont {Ansel}}, \bibinfo {author} {\bibfnamefont {E.}~\bibnamefont {Yang}}, \bibinfo {author} {\bibfnamefont {H.}~\bibnamefont {He}}, \bibinfo {author} {\bibfnamefont {N.}~\bibnamefont {Gimelshein}}, \bibinfo {author} {\bibfnamefont {A.}~\bibnamefont {Jain}}, \bibinfo {author} {\bibfnamefont {M.}~\bibnamefont {Voznesensky}}, \bibinfo {author} {\bibfnamefont {B.}~\bibnamefont {Bao}}, \bibinfo {author} {\bibfnamefont {P.}~\bibnamefont {Bell}}, \bibinfo {author} {\bibfnamefont {D.}~\bibnamefont {Berard}}, \bibinfo {author} {\bibfnamefont {E.}~\bibnamefont {Burovski}}, \bibinfo {author} {\bibfnamefont {G.}~\bibnamefont {Chauhan}}, \bibinfo {author} {\bibfnamefont {A.}~\bibnamefont {Chourdia}}, \bibinfo {author} {\bibfnamefont {W.}~\bibnamefont {Constable}}, \bibinfo {author} {\bibfnamefont {A.}~\bibnamefont {Desmaison}}, \bibinfo {author} {\bibfnamefont {Z.}~\bibnamefont {DeVito}}, \bibinfo {author} {\bibfnamefont {E.}~\bibnamefont {Ellison}},
  \bibinfo {author} {\bibfnamefont {W.}~\bibnamefont {Feng}}, \bibinfo {author} {\bibfnamefont {J.}~\bibnamefont {Gong}}, \bibinfo {author} {\bibfnamefont {M.}~\bibnamefont {Gschwind}}, \bibinfo {author} {\bibfnamefont {B.}~\bibnamefont {Hirsh}}, \bibinfo {author} {\bibfnamefont {S.}~\bibnamefont {Huang}}, \bibinfo {author} {\bibfnamefont {K.}~\bibnamefont {Kalambarkar}}, \bibinfo {author} {\bibfnamefont {L.}~\bibnamefont {Kirsch}}, \bibinfo {author} {\bibfnamefont {M.}~\bibnamefont {Lazos}}, \bibinfo {author} {\bibfnamefont {M.}~\bibnamefont {Lezcano}}, \bibinfo {author} {\bibfnamefont {Y.}~\bibnamefont {Liang}}, \bibinfo {author} {\bibfnamefont {J.}~\bibnamefont {Liang}}, \bibinfo {author} {\bibfnamefont {Y.}~\bibnamefont {Lu}}, \bibinfo {author} {\bibfnamefont {C.}~\bibnamefont {Luk}}, \bibinfo {author} {\bibfnamefont {B.}~\bibnamefont {Maher}}, \bibinfo {author} {\bibfnamefont {Y.}~\bibnamefont {Pan}}, \bibinfo {author} {\bibfnamefont {C.}~\bibnamefont {Puhrsch}}, \bibinfo {author} {\bibfnamefont
  {M.}~\bibnamefont {Reso}}, \bibinfo {author} {\bibfnamefont {M.}~\bibnamefont {Saroufim}}, \bibinfo {author} {\bibfnamefont {M.~Y.}\ \bibnamefont {Siraichi}}, \bibinfo {author} {\bibfnamefont {H.}~\bibnamefont {Suk}}, \bibinfo {author} {\bibfnamefont {M.}~\bibnamefont {Suo}}, \bibinfo {author} {\bibfnamefont {P.}~\bibnamefont {Tillet}}, \bibinfo {author} {\bibfnamefont {E.}~\bibnamefont {Wang}}, \bibinfo {author} {\bibfnamefont {X.}~\bibnamefont {Wang}}, \bibinfo {author} {\bibfnamefont {W.}~\bibnamefont {Wen}}, \bibinfo {author} {\bibfnamefont {S.}~\bibnamefont {Zhang}}, \bibinfo {author} {\bibfnamefont {X.}~\bibnamefont {Zhao}}, \bibinfo {author} {\bibfnamefont {K.}~\bibnamefont {Zhou}}, \bibinfo {author} {\bibfnamefont {R.}~\bibnamefont {Zou}}, \bibinfo {author} {\bibfnamefont {A.}~\bibnamefont {Mathews}}, \bibinfo {author} {\bibfnamefont {G.}~\bibnamefont {Chanan}}, \bibinfo {author} {\bibfnamefont {P.}~\bibnamefont {Wu}},\ and\ \bibinfo {author} {\bibfnamefont {S.}~\bibnamefont {Chintala}},\ }\bibfield
   {title} {\bibinfo {title} {{PyTorch 2: Faster Machine Learning Through Dynamic Python Bytecode Transformation and Graph Compilation}},\ }in\ \href {https://doi.org/10.1145/3620665.3640366} {\emph {\bibinfo {booktitle} {29th ACM International Conference on Architectural Support for Programming Languages and Operating Systems, Volume 2 (ASPLOS '24)}}}\ (\bibinfo  {publisher} {ACM},\ \bibinfo {year} {2024})\BibitemShut {NoStop}%
\bibitem [{\citenamefont {Seitz}\ \emph {et~al.}(2025)\citenamefont {Seitz}, \citenamefont {Heim}, \citenamefont {Moutinho}, \citenamefont {Guichard}, \citenamefont {Abramavicius}, \citenamefont {Wennersteen}, \citenamefont {Both}, \citenamefont {Quelle}, \citenamefont {Groot}, \citenamefont {Velikova}, \citenamefont {Elfving},\ and\ \citenamefont {Dagrada}}]{qadence2025}%
  \BibitemOpen
  \bibfield  {author} {\bibinfo {author} {\bibfnamefont {D.}~\bibnamefont {Seitz}}, \bibinfo {author} {\bibfnamefont {N.}~\bibnamefont {Heim}}, \bibinfo {author} {\bibfnamefont {J.}~\bibnamefont {Moutinho}}, \bibinfo {author} {\bibfnamefont {R.}~\bibnamefont {Guichard}}, \bibinfo {author} {\bibfnamefont {V.}~\bibnamefont {Abramavicius}}, \bibinfo {author} {\bibfnamefont {A.}~\bibnamefont {Wennersteen}}, \bibinfo {author} {\bibfnamefont {G.-J.}\ \bibnamefont {Both}}, \bibinfo {author} {\bibfnamefont {A.}~\bibnamefont {Quelle}}, \bibinfo {author} {\bibfnamefont {C.}~\bibnamefont {Groot}}, \bibinfo {author} {\bibfnamefont {G.}~\bibnamefont {Velikova}}, \bibinfo {author} {\bibfnamefont {V.}~\bibnamefont {Elfving}},\ and\ \bibinfo {author} {\bibfnamefont {M.}~\bibnamefont {Dagrada}},\ }\bibfield  {title} {\bibinfo {title} {Qadence: a differentiable interface for digital and analog programs},\ }\href {https://doi.org/10.1109/MS.2025.3536607} {\bibfield  {journal} {\bibinfo  {journal} {IEEE Software}\ }\textbf
  {\bibinfo {volume} {PP}},\ \bibinfo {pages} {1} (\bibinfo {year} {2025})}\BibitemShut {NoStop}%
\bibitem [{\citenamefont {Zhang}\ \emph {et~al.}(2017)\citenamefont {Zhang}, \citenamefont {Bengio}, \citenamefont {Hardt}, \citenamefont {Recht},\ and\ \citenamefont {Vinyals}}]{zhang2017}%
  \BibitemOpen
  \bibfield  {author} {\bibinfo {author} {\bibfnamefont {C.}~\bibnamefont {Zhang}}, \bibinfo {author} {\bibfnamefont {S.}~\bibnamefont {Bengio}}, \bibinfo {author} {\bibfnamefont {M.}~\bibnamefont {Hardt}}, \bibinfo {author} {\bibfnamefont {B.}~\bibnamefont {Recht}},\ and\ \bibinfo {author} {\bibfnamefont {O.}~\bibnamefont {Vinyals}},\ }\bibfield  {title} {\bibinfo {title} {Understanding deep learning requires rethinking generalization},\ }in\ \href@noop {} {\emph {\bibinfo {booktitle} {International Conference on Learning Representations (ICLR)}}}\ (\bibinfo {year} {2017})\BibitemShut {NoStop}%
\bibitem [{\citenamefont {Neyshabur}\ \emph {et~al.}(2015)\citenamefont {Neyshabur}, \citenamefont {Tomioka},\ and\ \citenamefont {Srebro}}]{neyshabur2015}%
  \BibitemOpen
  \bibfield  {author} {\bibinfo {author} {\bibfnamefont {B.}~\bibnamefont {Neyshabur}}, \bibinfo {author} {\bibfnamefont {R.}~\bibnamefont {Tomioka}},\ and\ \bibinfo {author} {\bibfnamefont {N.}~\bibnamefont {Srebro}},\ }\bibfield  {title} {\bibinfo {title} {In search of the real inductive bias: On the role of implicit regularization in deep learning},\ }in\ \href@noop {} {\emph {\bibinfo {booktitle} {International Conference on Learning Representations (ICLR) Workshop}}}\ (\bibinfo {year} {2015})\BibitemShut {NoStop}%
\bibitem [{\citenamefont {Anschuetz}(2022)}]{anschuetz_2021_critical}%
  \BibitemOpen
  \bibfield  {author} {\bibinfo {author} {\bibfnamefont {E.~R.}\ \bibnamefont {Anschuetz}},\ }\bibfield  {title} {\bibinfo {title} {Critical points in quantum generative models},\ }in\ \href {https://openreview.net/forum?id=2f1z55GVQN} {\emph {\bibinfo {booktitle} {International Conference on Learning Representations (ICLR)}}}\ (\bibinfo {year} {2022})\BibitemShut {NoStop}%
\bibitem [{\citenamefont {Bittel}\ and\ \citenamefont {Kliesch}(2021)}]{bittel_2021_training}%
  \BibitemOpen
  \bibfield  {author} {\bibinfo {author} {\bibfnamefont {L.}~\bibnamefont {Bittel}}\ and\ \bibinfo {author} {\bibfnamefont {M.}~\bibnamefont {Kliesch}},\ }\bibfield  {title} {\bibinfo {title} {Training variational quantum algorithms is np-hard},\ }\href {https://doi.org/10.1103/PhysRevLett.127.120502} {\bibfield  {journal} {\bibinfo  {journal} {Phys. Rev. Lett.}\ }\textbf {\bibinfo {volume} {127}},\ \bibinfo {pages} {120502} (\bibinfo {year} {2021})}\BibitemShut {NoStop}%
\bibitem [{\citenamefont {Pesah}\ \emph {et~al.}(2021)\citenamefont {Pesah}, \citenamefont {Cerezo}, \citenamefont {Wang}, \citenamefont {Volkoff}, \citenamefont {Sornborger},\ and\ \citenamefont {Coles}}]{pesah_physrevX_2021}%
  \BibitemOpen
  \bibfield  {author} {\bibinfo {author} {\bibfnamefont {A.}~\bibnamefont {Pesah}}, \bibinfo {author} {\bibfnamefont {M.}~\bibnamefont {Cerezo}}, \bibinfo {author} {\bibfnamefont {S.}~\bibnamefont {Wang}}, \bibinfo {author} {\bibfnamefont {T.}~\bibnamefont {Volkoff}}, \bibinfo {author} {\bibfnamefont {A.~T.}\ \bibnamefont {Sornborger}},\ and\ \bibinfo {author} {\bibfnamefont {P.~J.}\ \bibnamefont {Coles}},\ }\bibfield  {title} {\bibinfo {title} {Absence of barren plateaus in quantum convolutional neural networks},\ }\href {https://doi.org/10.1103/PhysRevX.11.041011} {\bibfield  {journal} {\bibinfo  {journal} {Phys. Rev. X}\ }\textbf {\bibinfo {volume} {11}},\ \bibinfo {pages} {041011} (\bibinfo {year} {2021})}\BibitemShut {NoStop}%
\bibitem [{\citenamefont {Cerezo}\ \emph {et~al.}(2021)\citenamefont {Cerezo}, \citenamefont {Sone}, \citenamefont {Volkoff}, \citenamefont {Cincio},\ and\ \citenamefont {Coles}}]{cerezo_nature_2021_costfunc}%
  \BibitemOpen
  \bibfield  {author} {\bibinfo {author} {\bibfnamefont {M.}~\bibnamefont {Cerezo}}, \bibinfo {author} {\bibfnamefont {A.}~\bibnamefont {Sone}}, \bibinfo {author} {\bibfnamefont {T.}~\bibnamefont {Volkoff}}, \bibinfo {author} {\bibfnamefont {L.}~\bibnamefont {Cincio}},\ and\ \bibinfo {author} {\bibfnamefont {P.~J.}\ \bibnamefont {Coles}},\ }\bibfield  {title} {\bibinfo {title} {Cost function dependent barren plateaus in shallow parametrized quantum circuits},\ }\href {https://doi.org/doi.org/10.1038/s41467-021-21728-w} {\bibfield  {journal} {\bibinfo  {journal} {Nat Commun}\ }\textbf {\bibinfo {volume} {12}},\ \bibinfo {pages} {1791} (\bibinfo {year} {2021})}\BibitemShut {NoStop}%
\bibitem [{\citenamefont {Cerezo}\ \emph {et~al.}(2025)\citenamefont {Cerezo}, \citenamefont {Larocca}, \citenamefont {García-Martín}, \citenamefont {Diaz}, \citenamefont {Braccia}, \citenamefont {Fontana}, \citenamefont {Rudolph}, \citenamefont {Bermejo}, \citenamefont {Ijaz}, \citenamefont {Thanasilp}, \citenamefont {Anschuetz},\ and\ \citenamefont {Holmes}}]{cerezo_provable_absence_2024}%
  \BibitemOpen
  \bibfield  {author} {\bibinfo {author} {\bibfnamefont {M.}~\bibnamefont {Cerezo}}, \bibinfo {author} {\bibfnamefont {M.}~\bibnamefont {Larocca}}, \bibinfo {author} {\bibfnamefont {D.}~\bibnamefont {García-Martín}}, \bibinfo {author} {\bibfnamefont {N.~L.}\ \bibnamefont {Diaz}}, \bibinfo {author} {\bibfnamefont {P.}~\bibnamefont {Braccia}}, \bibinfo {author} {\bibfnamefont {E.}~\bibnamefont {Fontana}}, \bibinfo {author} {\bibfnamefont {M.~S.}\ \bibnamefont {Rudolph}}, \bibinfo {author} {\bibfnamefont {P.}~\bibnamefont {Bermejo}}, \bibinfo {author} {\bibfnamefont {A.}~\bibnamefont {Ijaz}}, \bibinfo {author} {\bibfnamefont {S.}~\bibnamefont {Thanasilp}}, \bibinfo {author} {\bibfnamefont {E.~R.}\ \bibnamefont {Anschuetz}},\ and\ \bibinfo {author} {\bibfnamefont {Z.}~\bibnamefont {Holmes}},\ }\bibfield  {title} {\bibinfo {title} {Does provable absence of barren plateaus imply classical simulability?},\ }\href {https://doi.org/10.1038/s41467-025-63099-6} {\bibfield  {journal} {\bibinfo  {journal} {Nat Commun}\
  }\textbf {\bibinfo {volume} {16}},\ \bibinfo {pages} {7907} (\bibinfo {year} {2025})}\BibitemShut {NoStop}%
\bibitem [{\citenamefont {Cerezo}\ and\ \citenamefont {Coles}(2021)}]{cerezo_2020_impact}%
  \BibitemOpen
  \bibfield  {author} {\bibinfo {author} {\bibfnamefont {M.}~\bibnamefont {Cerezo}}\ and\ \bibinfo {author} {\bibfnamefont {P.~J.}\ \bibnamefont {Coles}},\ }\bibfield  {title} {\bibinfo {title} {Higher order derivatives of quantum neural networks with barren plateaus},\ }\href {https://doi.org/10.1088/2058-9565/abf51a} {\bibfield  {journal} {\bibinfo  {journal} {Quantum Science and Technology}\ }\textbf {\bibinfo {volume} {6}},\ \bibinfo {pages} {035006} (\bibinfo {year} {2021})}\BibitemShut {NoStop}%
\bibitem [{\citenamefont {McClean}\ \emph {et~al.}(2018)\citenamefont {McClean}, \citenamefont {Boixo}, \citenamefont {Smelyanskiy}, \citenamefont {Babbush},\ and\ \citenamefont {Neven}}]{mcclean_2018_barren}%
  \BibitemOpen
  \bibfield  {author} {\bibinfo {author} {\bibfnamefont {J.~R.}\ \bibnamefont {McClean}}, \bibinfo {author} {\bibfnamefont {S.}~\bibnamefont {Boixo}}, \bibinfo {author} {\bibfnamefont {V.~N.}\ \bibnamefont {Smelyanskiy}}, \bibinfo {author} {\bibfnamefont {R.}~\bibnamefont {Babbush}},\ and\ \bibinfo {author} {\bibfnamefont {H.}~\bibnamefont {Neven}},\ }\bibfield  {title} {\bibinfo {title} {Barren plateaus in quantum neural network training landscapes},\ }\href {https://www.nature.com/articles/s41467-018-07090-4} {\bibfield  {journal} {\bibinfo  {journal} {Nat Commun}\ }\textbf {\bibinfo {volume} {9}},\ \bibinfo {pages} {4812} (\bibinfo {year} {2018})}\BibitemShut {NoStop}%
\bibitem [{\citenamefont {Holmes}\ \emph {et~al.}(2022)\citenamefont {Holmes}, \citenamefont {Sharma}, \citenamefont {Cerezo},\ and\ \citenamefont {Coles}}]{holmes_2021_connecting}%
  \BibitemOpen
  \bibfield  {author} {\bibinfo {author} {\bibfnamefont {Z.}~\bibnamefont {Holmes}}, \bibinfo {author} {\bibfnamefont {K.}~\bibnamefont {Sharma}}, \bibinfo {author} {\bibfnamefont {M.}~\bibnamefont {Cerezo}},\ and\ \bibinfo {author} {\bibfnamefont {P.~J.}\ \bibnamefont {Coles}},\ }\bibfield  {title} {\bibinfo {title} {Connecting ansatz expressibility to gradient magnitudes and barren plateaus},\ }\href {https://doi.org/10.1103/PRXQuantum.3.010313} {\bibfield  {journal} {\bibinfo  {journal} {PRX Quantum}\ }\textbf {\bibinfo {volume} {3}},\ \bibinfo {pages} {010313} (\bibinfo {year} {2022})}\BibitemShut {NoStop}%
\bibitem [{\citenamefont {Bermejo}\ \emph {et~al.}(2024)\citenamefont {Bermejo}, \citenamefont {Aizpurua},\ and\ \citenamefont {Or\'us}}]{bermejo_2024_improving}%
  \BibitemOpen
  \bibfield  {author} {\bibinfo {author} {\bibfnamefont {P.}~\bibnamefont {Bermejo}}, \bibinfo {author} {\bibfnamefont {B.}~\bibnamefont {Aizpurua}},\ and\ \bibinfo {author} {\bibfnamefont {R.}~\bibnamefont {Or\'us}},\ }\bibfield  {title} {\bibinfo {title} {Improving gradient methods via coordinate transformations: Applications to quantum machine learning},\ }\href {https://doi.org/10.1103/PhysRevResearch.6.023069} {\bibfield  {journal} {\bibinfo  {journal} {Phys. Rev. Res.}\ }\textbf {\bibinfo {volume} {6}},\ \bibinfo {pages} {023069} (\bibinfo {year} {2024})}\BibitemShut {NoStop}%
\end{thebibliography}%
\appendix
\onecolumngrid   

\section{Numerical Test of Convergence Speed Under Different Initializations}
\label{app:seed_robustness}

\begin{figure}[H]
    \centering
    \begin{subfigure}{0.49\columnwidth}
        \centering
        \includegraphics[width=\linewidth]{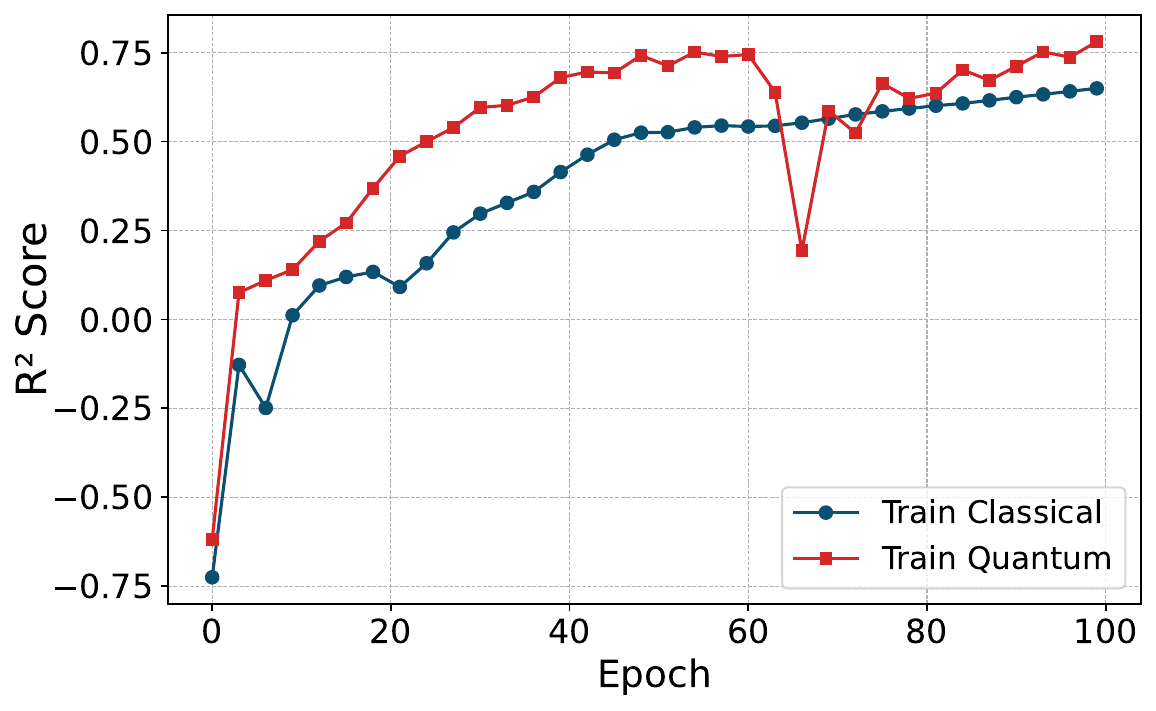}
        \caption{Training R$^2$ score (seed 35).}
    \end{subfigure}
    \hfill
    \begin{subfigure}{0.49\columnwidth}
        \centering
        \includegraphics[width=\linewidth]{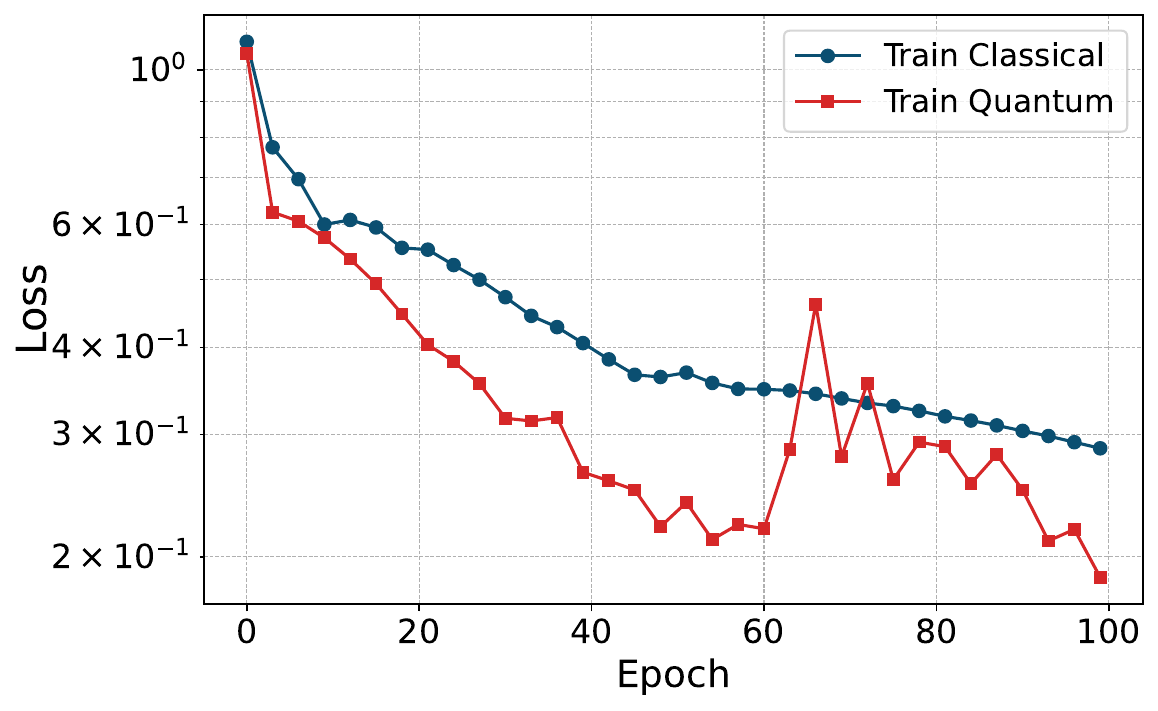}
        \caption{Training loss (seed 35).}
    \end{subfigure}
    \begin{subfigure}{0.49\columnwidth}
        \centering
        \includegraphics[width=\linewidth]{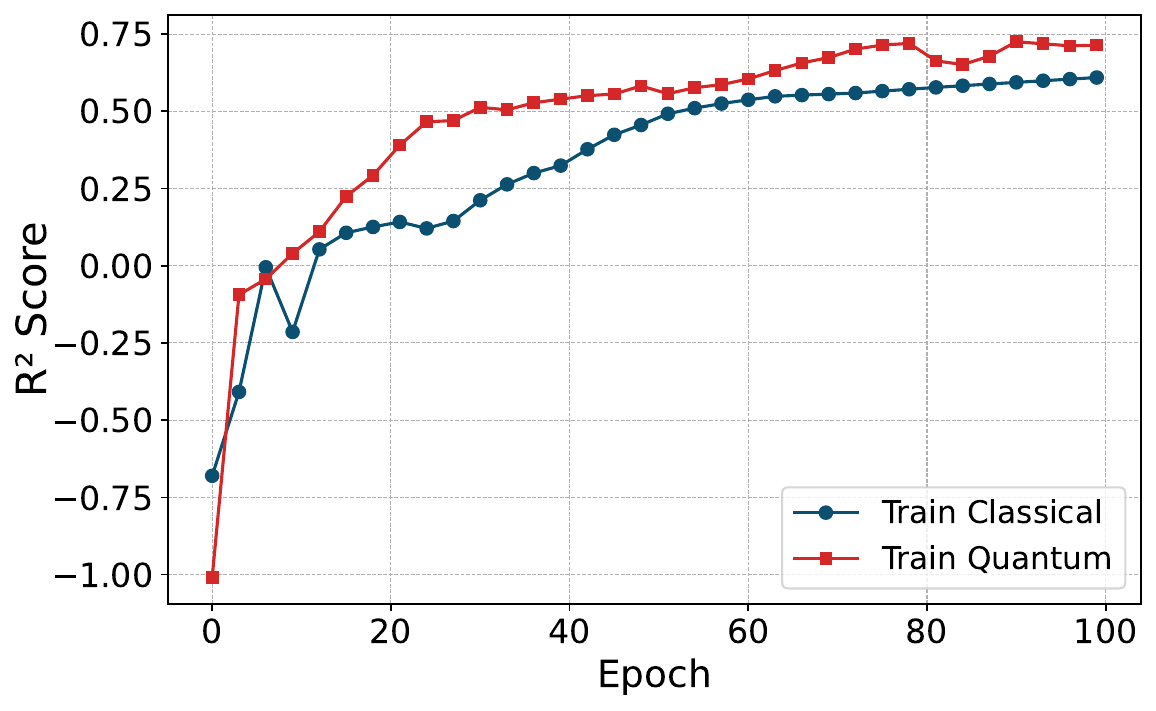}
        \caption{Training R$^2$ score (seed 42).}
    \end{subfigure}
    \hfill
    \begin{subfigure}{0.49\columnwidth}
        \centering
        \includegraphics[width=\linewidth]{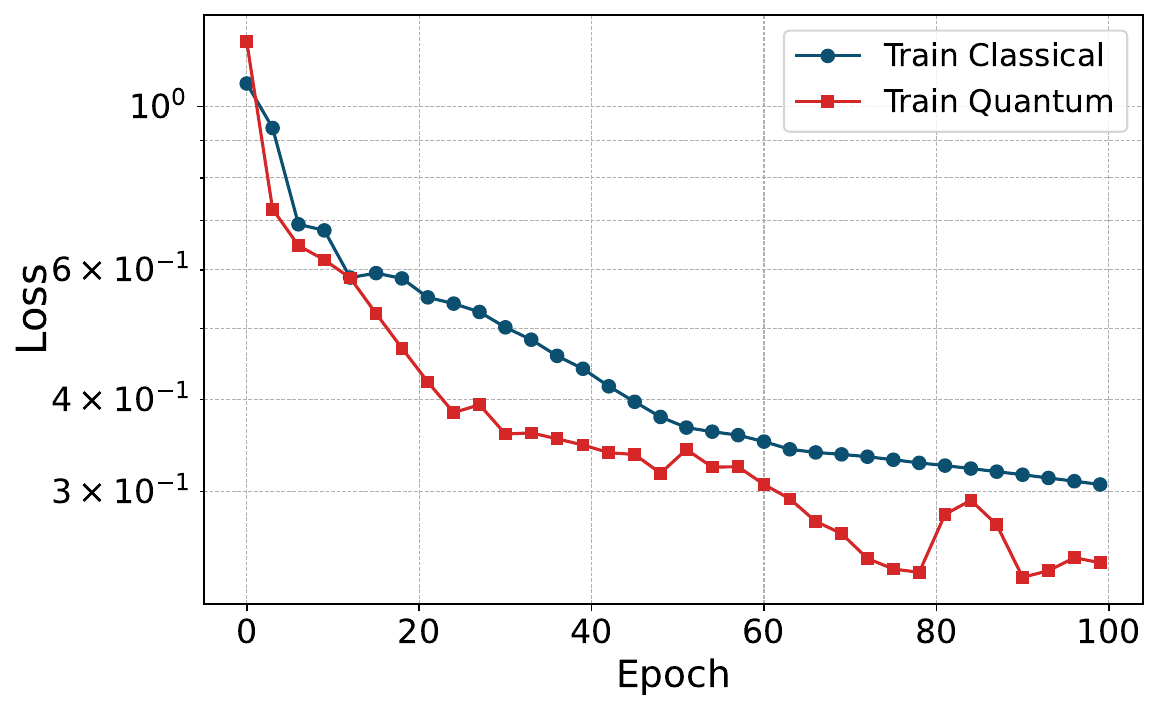}
        \caption{Training loss (seed 42).}
    \end{subfigure}
    \caption{Training R$^2$ score and loss over the first 100 epochs for a sample of 30 molecules with up to 14 atoms, under seed 35 (top) and seed 42 (bottom). The QGNN (red squares) shows faster early-training convergence than the classical GNN (blue circles) under both seeds, indicating that this behavior is not an artifact of the particular random seed used.}
    \label{fig:seed}
\end{figure}

In this appendix, we assess whether the faster initial learning observed for the QGNN relative to the classical GNN is an artifact of a favorable parameter initialization rather than a robust property of the quantum aggregator. To this end, we consider a sample of 30 molecules with up to 14 atoms from the QM9 dataset, a molecule range distinct from those considered in the main text, to confirm that the effect does not depend on the particular sample selection used there. Now, we consider two random seeds, $35$ and $42$, while keeping all other hyperparameters identical to Case~2 in the main text (Table~ \ref{tab:comparison}): a GNN with hidden-layer architecture $[8,4]$ and $305$ trainable parameters, and a QGNN with a single circuit of depth configuration $r=[3,3,3]$ and $293$ trainable parameters and the same Fourier feature map encoding, both trained with an initial learning rate of $0.03$. Figure~\ref{fig:seed} shows the training R$^2$ score and loss over the first 100 epochs for both seeds. In both cases, the QGNN reaches higher R$^2$ scores and lower loss earlier in training than the classical GNN, confirming that the faster initial learning of the quantum model persists across different random seeds and molecule samples, and is therefore not an artifact of a particular favorable initialization.

\section{Effect of Feature Encoding on Model Performance}
\label{app:fourier_ablation}

In this appendix, we assess whether the performance advantage of the QGNN originates from the Fourier feature map encoding rather than from the variational quantum circuit itself. To this end, we consider a sample of 30 molecules with up to $18$ atoms from the QM9 dataset, with $70$\% of the sample used for training and $30$\% for testing, trained over 300 epochs. For the QGNN baseline, we use a quantum circuit with depth configuration $r = [3,3,3]$ and $293$ trainable parameters, while the classical GNN uses a GraphSAGE-style architecture with a single MLP and hidden-layer configuration of $[8,4]$ and $305$ trainable parameters. Both models are trained with an initial learning rate of $0.03$, identical to Case 2 of the main text (Table~\ref{tab:comparison}).

\begin{figure}[H]
    \centering
    \begin{subfigure}{0.49\columnwidth}
        \centering
        \includegraphics[width=\linewidth]{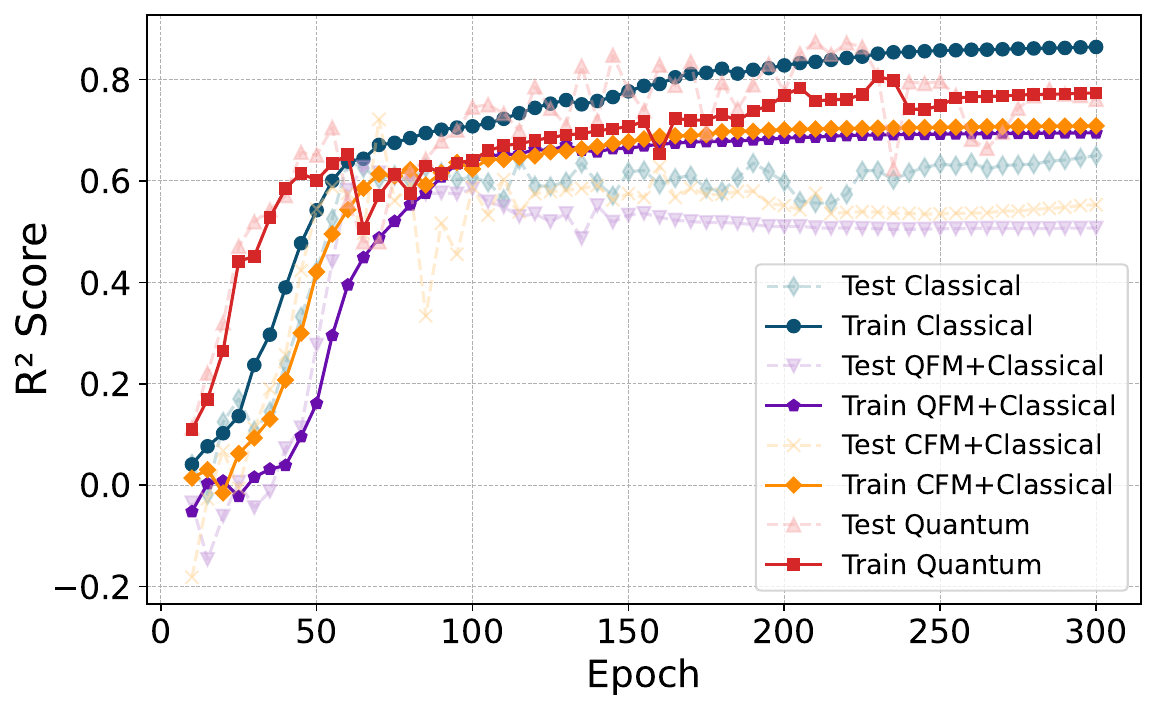}
        \caption{R$^2$ score curves}
        \label{fig:case2_score}
    \end{subfigure}
    \hfill
    \begin{subfigure}{0.49\columnwidth}
        \centering
        \includegraphics[width=\linewidth]{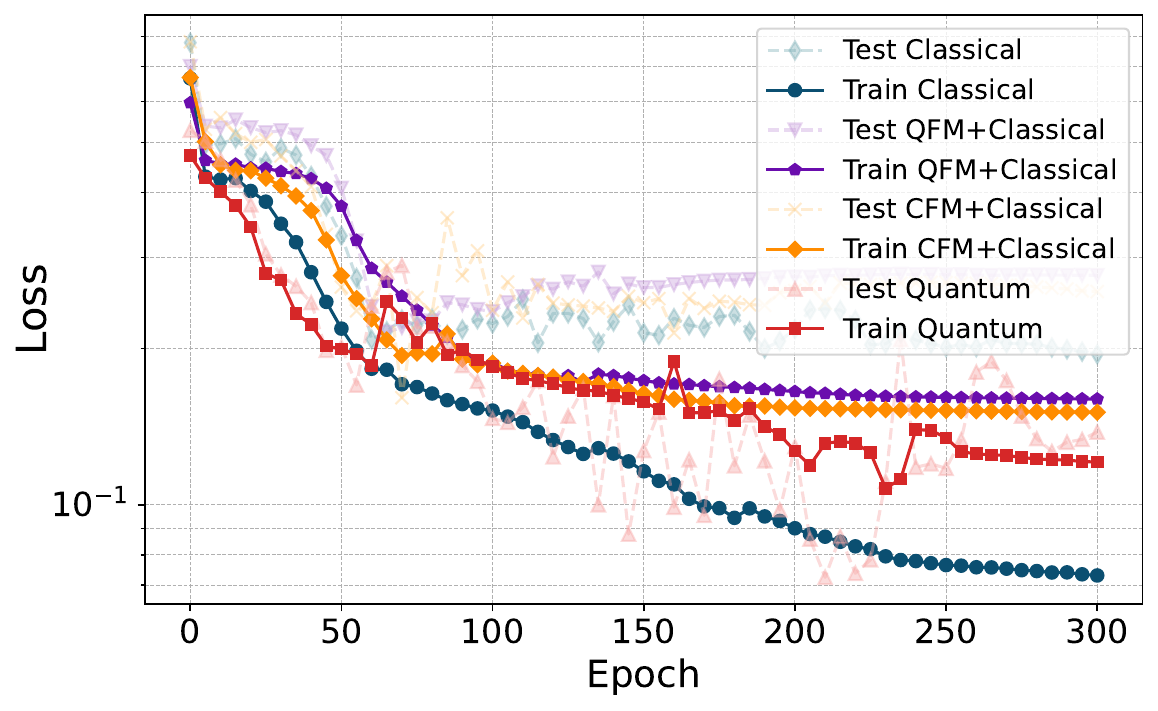}
        \caption{Loss curves.}
        \label{fig:case2_loss}
    \end{subfigure}
    \caption{R$^2$ scores (a) and losses (b) over 300 training epochs for the standard classical GNN, standard QGNN, QFM-augmented GNN, and CFM-augmented GNN. Both augmented baselines underperform relative to the standard models.}
    \label{fig:case2_ablation}
\end{figure}

We introduce two non-trainable encoding baselines, each feeding a transformed feature vector into an otherwise standard classical GNN: a Quantum Fourier Map (QFM), using the same single-qubit Pauli-rotation feature map as in~\cite{schuld_2021} with a fixed observable measurement in place of a variational circuit, and a Classical Fourier Map (CFM), computing $\cos(x_i)$ directly as a classical analogue of the same encoding. As shown in Fig.~ \ref{fig:case2_ablation}, both augmented baselines underperform relative to the standard classical GNN and the standard QGNN alike, rather than closing the observed performance gap. This suggests that the performance advantage of the QGNN, in particular its generalization advantage, cannot be attributed to the feature encoding alone, and instead points to the end-to-end trained quantum aggregation mechanism as the primary source of the observed gains.

\vspace{2mm}

\end{document}